\documentclass[pra,aps,amssymb,twocolumn,superscriptaddress, notitlepage,longbibliography]{revtex4-2}

\usepackage{mathtools}
\usepackage{graphicx} 
\usepackage{xcolor}
\usepackage{float}
\usepackage[colorlinks = true,
            linkcolor = blue,
            urlcolor  = blue,
            citecolor = blue,
            anchorcolor = blue]{hyperref}
\usepackage[normalem]{ulem}
\usepackage{physics}
\usepackage{orcidlink}
\usepackage{appendix}
\usepackage{amsfonts}

\usepackage{pgfplots}
\usepackage{amsmath}
\usepackage{comment}
\usepackage{ifthen}
\usepackage{tikz-3dplot}
\usetikzlibrary{calc}
\usetikzlibrary{intersections,shapes.arrows}
\pgfplotsset{compat=1.17}

\tdplotsetmaincoords{60}{30}

\newcommand{\sect}[1]{\emph{#1}---}

\newcommand{\unilu}{Department of Physics and Materials Science, University of Luxembourg, L-1511 Luxembourg, Luxembourg}

\newcommand{\moyb}[2]{\{\{#1,#2\}\}}
\newcommand{\moybst}[2]{\{\{#1,#2\}\}_{+}}
\begin{document}      

\title{Double-Bracket Master Equations: Phase-Space Representation and Classical Limit}


\author{Ankit W. Shrestha \orcidlink{0009-0008-5367-7062}}
    \email{ankit.wenjushrestha@uni.lu} 
    \affiliation{\unilu}

\author{Budhaditya Bhattacharjee \orcidlink{0000-0003-1982-1346}}
\email{budhaditya.bhattacharjee@uni.lu}
    \affiliation{\unilu}

\author{Adolfo~del Campo\orcidlink{0000-0003-2219-2851}} 
    \email{adolfo.delcampo@uni.lu}
    \affiliation{\unilu}
    \affiliation{Donostia International Physics Center, E-20018 San Sebastián, Spain}

\date{\today}

\begin{abstract}
We systematically investigate master equations involving double-bracket dissipators in the classical limit. Dissipators defined by double commutators arise naturally in dephasing dynamics, while those defined by double anticommutators are typically associated with noisy Hamiltonians. The classical limit of such master equations is derived by reformulating the dissipative open dynamics in phase space using the Wigner–Weyl transform and Moyal bracket formalism, and by identifying the leading-order terms in a systematic $\hbar$-expansion.
We first analyze a double-commutator master equation associated with energy diffusion, and then turn to master equations containing a double anticommutator with the system Hamiltonian, recently derived in the context of noisy non-Hermitian systems. For both classes of double-bracket equations, we establish a gradient-flow representation of the dynamics within both the quantum and semiclassical formalisms.
We further study the semiclassical evolution generated by double-bracket master equations for the harmonic (integrable) and driven anharmonic (chaotic) oscillators, considering both classical and quantum initial states. The dynamics are characterized through several observables, including mean position, momentum, and energy. The quantumness of the time-evolved state is assessed via the Wigner logarithmic negativity, and the interplay between double-bracket dissipation and chaotic dynamics is examined in detail.
Finally, we extend our analysis to generalized master equations involving higher-order nested brackets, which arise naturally as a time-continuous formulation of spectral filtering techniques widely used in the numerical simulation of quantum systems.
\end{abstract}
\maketitle

\section{Introduction}
Double-bracket master equations govern the evolution of physical systems in a variety of contexts. A familiar instance arises in the open quantum dynamics of Markovian systems \cite{BP02}. The associated master equation takes the Lindblad form \cite{Lindblad76}, and when it involves a single Hermitian jump operator, it reduces to a double-commutator master equation that describes dephasing in the eigenbasis of the jump operator.  
When the latter coincides with the system Hamiltonian, the evolution describes energy dephasing,
which constitutes a simple model of decoherence with manifold applications. 
An ubiquitous scenario in which it arises involves the description of unitary evolutions timed with realistic clocks involving errors \cite{egusquiza1999quantum,egusquiza2003real}. Similarly, it can be derived as an effective dissipative dynamics describing an ensemble of trajectories governed by stochastic Hamiltonians \cite{chenu2017quantum}. In this context, it has been used to predict extreme decoherence in noisy random Hamiltonians \cite{xu2019extreme} and to describe the emergence of objectivity from random quantum measurements \cite{korbicz2017generic}. Energy dephasing models also allow us to describe the use of filters in the numerical study of spectral statistics in many-body systems as physical operations characterized by a quantum channel \cite{apollo2023unitarity}. As a result, it provides a natural framework for investigating dissipative quantum chaos \cite{Xu21SFF,MatsoukasRoubeas2024} and the role of decoherence in conformal field theory and holography \cite{delCampo2020}. Energy dephasing also arises in modifications of quantum mechanics of the kind involved in wavefunction collapse models \cite{gisin1984quantum,percival1994primary,schneider1998decoherence,Adler01,Brody02,adler2003weisskopf,bassi2003dynamical,Brody06,bassi2013models}. In addition, it can be seen as a Liouvillian deformation of Hamiltonian evolution \cite{MatsoukasRoubeas23}.

It is worth noting that the kind of double-bracket master equation that describes energy dephasing arises in a broader context related to matrix diagonalization \cite{Brockett91}. For example, the Hamiltonian flow formulated by Wegner \cite{Wegner94,Wegner2001} and Glazek and Wilson \cite{GlazekWilson93,GlazekWilson94} in the context of the continuous renormalization group is governed by a double-bracket equation that diagonalizes the Hamiltonian in block form \cite{Kehrein2007}. The analogy of the Wegner flow with energy dephasing has been discussed in \cite{Hornedal23}. This kind of double-bracket master equation is at the core of a variety of quantum algorithms \cite{Gluza24,Gluza26}. 

More recently, double bracket equations have been introduced in the context of 
noisy non-Hermitian evolutions. In this case, the equation of motion involves a double anticommutator with the imaginary part  of the fluctuating operator, such as the system Hamiltonian \cite{pablo2025quantum}. 

Double-bracket equations have numerous applications, and it is natural to explore their classical limit. In this work, we undertake such an analysis by resorting to a phase space formulation of the quantum evolution. We analyze the dissipative time-evolution of the Wigner function and perform an $\hbar$-expansion to identify the corresponding classical equations of motion. 

We first discuss how master equations with a double-commutator arise in a variety of contexts in Sec. \ref{SecDCME}.  By formulating the master equations in phase space, we discuss their classical limit in Sec.  \ref{SecDCMEcl}. The analysis is generalized to master equations involving double anticommutators in Sec. \ref{SecDAME}. The gradient flow interpretation of master equations involving either kind of double brackets is presented in Sec. \ref{SecGradfow}. Section \ref{SecExamples} illustrates the evolution under double-bracket master equations in the simple case of a quantum oscillator, characterizing natural observables and the Wigner negativity. A driven anharmonic oscillator is studied to explore the interplay of decoherence in the classical-quantum correspondence of chaotic models. In addition, we discuss how filters used in the numerical analysis of spectral properties of quantum systems can be described by generalized master equations involving higher-order nested brackets, in Sec \ref{SecGenME}, before closing with a summary and discussion.

\section{Markovian Quantum Dynamics And Double-Commutator Master evolutions} \label{SecDCME}

When the state of a system embedded in a surrounding environment is initially described by a tensor product state, a Markovian master equation for the reduced state of the system can be derived. The latter is known as the Gorini–Kossakowski–Sudarshan–Lindblad (GKSL) master equation and can be written in terms of the Hamiltonian of the system (including the Lamb shift) and dissipation terms \cite{Lindblad76,GKS76,BP02}
\begin{equation}
    \dot{\hat{\rho}} = -\frac{i}{\hbar}[\hat{H},\hat{\rho}] + \sum_{m}\Gamma_m\mathcal{D}[\hat{L}_m]\hat{\rho}\,,
\end{equation}
where $\mathcal{D}[\hat{L}_m]\hat{\rho} = \hat{L}^\dagger_m \hat{\rho} \hat{L}_m - \frac{1}{2}\{\hat{L}^\dagger_m \hat{L}_m,\hat{\rho}\}$ is known as the dissipator. This evolution arises in a large class of quantum systems.
For Hermitian jump operators, the quantum GKSL master equation reduces to a double-bracket equation 
\begin{equation}
\dot{\hat{\rho}} = -\frac{i}{\hbar}[\hat{H},\hat{\rho}] - \frac{1}{2}\sum_{m}\Gamma_m[\hat{L}_m,[\hat{L}_m,\hat{\rho}]]\,. \label{DMEQ}
\end{equation}
This evolution describes dephasing in the eigenbasis of $\hat{L}_m$. For instance, the high-temperature limit of quantum Brownian motion corresponds to the choice of the position operator as the single jump operator \cite{BP02,Zurek03}. The case of multiple noncommuting jump operators is also frequent and can be used to describe, e.g.,  the depolarizing channel of a qubit \cite{nielsen2010quantum} or phase-space measurements \cite{Brody25}.

The dephasing master equation can also be derived in a different context, that of fluctuating Hamiltonians. Fluctuating Hamiltonians have broad applications in physics and chemistry \cite{VanKampen92}, and can be used, e.g., to explore the robustness to noise of quantum control protocols such as shortcuts to adiabaticity \cite{Ruschhaupt2012,Kiely2017,delCampo2019,Kiely2021}, describing errors in quantum annealing \cite{Dutta16,Ai2021,Iwamura24}, and the quantum simulation of open systems \cite{chenu2017quantum,Smith2018}. Consider an isolated system, under unitary dynamics generated by the stochastic Hermitian Hamiltonian
\begin{equation}
\hat{H}_{\rm st}=\hat{H}+\sum_{m}\xi_{m}(t)\lambda_m\hat{L}_m\,,
\end{equation}
where $\lambda_m$ is a dimensionless constant and $\{\xi_{m}(t)\}$ represent real Gaussian processes with zero mean and white-noise autocorrelations $\langle\xi_m(t))\xi_n(t')\rangle=\delta_{mn}\delta(t-t')$.   
The evolution of an initial state $|\psi_{\rm st}\rangle$ is then dictated by the stochastic Schr\"odinger equation (or the Liouville von Neumann equation in the case of density matrices). Alternatively, one can consider an ensemble of realizations and introduce the noise-averaged quantum state $\hat{\rho}=\mathbb{E}[|\psi_{\rm st}\rangle\langle \psi_{\rm st}|]$, which can be shown to evolve according to the dephasing master equation (\ref{DMEQ}) with the identification \cite{VanKampen92,Budini00,Budini01,chenu2017quantum}
\begin{eqnarray}
\Gamma_m=(\lambda_m/\hbar)^2\,. \label{Gmeq}
\end{eqnarray}
Yet a different context for the occurrence of the dephasing master equation arises in the description of continuous quantum measurements \cite{Jacobs2006,Jacobs2014}. The monitoring of a set of observables $\hat{L}_m$ is described by a stochastic master equation. Disregarding the measurement outcomes, the state of the system is described by the dephasing master equation (\ref{DMEQ}).

The evolution (\ref{DMEQ}) simplifies when the dissipator takes the form of a single double commutator involving the Hamiltonian, i.e., by choosing the Hamiltonian as the only Hermitian Lindblad operator $\hat{L}=\hat{L}^\dag=\hat{H}$. The master equation for energy dephasing is given by
\begin{align}
    \dot{\hat{\rho}} = -\frac{i}{\hbar}[\hat{H},\hat{\rho}] - \frac{\Gamma}{2} [\hat{H},[\hat{H},\hat{\rho}]]\,, \label{ME4ED}
\end{align}
where $\Gamma$ indicates the strength of decoherence.  This evolution induces dephasing in the energy eigenbasis but preserves the energy distribution. As already mentioned, energy dephasing arises in a variety of scenarios, such as evolutions timed with realistic clocks,  fluctuating Hamiltonians and random measurements \cite{egusquiza1999quantum,egusquiza2003real,korbicz2017generic}, modifications of quantum mechanics \cite{gisin1984quantum,percival1994primary,schneider1998decoherence,adler2003weisskopf,bassi2003dynamical,bassi2013models}, spectral filtering \cite{MatsoukasRoubeas23,apollo2023unitarity}, and quantum chaos \cite{xu2019extreme,Xu21SFF}.

\section{Phase-space representation of Markovian open quantum evolutions and their Classical Limit
} \label{SecDCMEcl}

One way of formulating a classical limit is by resorting to a phase-space formulation in which the master equation can be understood as a $\hbar$-deformation of the classical dynamics, associated with the $\hbar \rightarrow 0$ limit \cite{Wigner32,Moyal1949quantum,Hillery84,Zachos05}. 
Consider a $N$-dimensional system with coordinates $\mathbf{x}=(x_1,\cdots,x_N)$, $\mathbf{p}=(p_1,\cdots,p_N)$, and let ${\rm d}\mathbf{x}={\rm d}x_1\cdots{\rm d}x_N$.  
The role of the density matrix is replaced by the Wigner function
\begin{equation}
W(\mathbf{x},\mathbf{p})=\frac{1}{(2\pi \hbar)^N}\int {\rm d}\mathbf{y}\langle {\bf x}-{\bf y}/2|\hat{\rho}|{\bf x}+{\bf y}/2\rangle e^{i{\bf p}\cdot{\bf y}/\hbar}\,,
\end{equation}
and to each operator $\hat{A}$ one associates the Weyl symbol $A=A(\mathbf{x},\mathbf{p})$
\begin{equation}
A(\mathbf{x},\mathbf{p})=\int  {\rm d}\mathbf{y}\langle {\bf x}-{\bf y}/2|\hat{A}|{\bf x}+{\bf y}/2\rangle e^{i{\bf p}\cdot{\bf y}/\hbar}\, .\label{DefWeyl}
\end{equation}
The expectation values can then be computed as
\begin{equation}
\langle \hat{A}\rangle={\rm Tr}(\hat{A}\hat{\rho})=\int{\rm d}\mathbf{x}{\rm d}\mathbf{p}\,A(\mathbf{x},\mathbf{p})W(\mathbf{x},\mathbf{p})\,, \label{avg A}
\end{equation}
and thus $\int{\rm d}\mathbf{x}{\rm d}\mathbf{p}W(\mathbf{x},\mathbf{p})=1$. 
We define the phase space average of a Weyl symbol $A$ as $\langle A\rangle =\int{\rm d}\mathbf{x}{\rm d}\mathbf{p}\,A(\mathbf{x},\mathbf{p})W(\mathbf{x},\mathbf{p})$.

Similarly, one can compute moments $\hat{A}^n$ of an observable as
\begin{equation}
\langle \hat{A}^n\rangle={\rm Tr}(\hat{A}\hat{\rho})=\langle A\star A \star\cdots\star A\rangle=\langle A^{\star n}\rangle\,,\label{avgAn}
\end{equation}
in terms of the star product
\begin{align}
A \star B &= A \exp\left\{\frac{i\hbar}{2}\sum_{j = 1}^{N}\left(\overleftarrow{\partial}_{x_j}\overrightarrow{\partial}_{p_j} - \overleftarrow{\partial}_{p_j}\overrightarrow{\partial}_{x_j}\right)\right\} B \notag\\ &= \sum_{j = 1}^{N}\sum_{s = 0}^{\infty}\frac{1}{s!}\left(\frac{i \hbar}{2}\right)^s \sum_{t = 0}^{s}(-1)^t \binom{s}{t}(\partial^{s-t}_{x_j}\partial^{t}_{p_j}A)\notag\\&\times (\partial^{t}_{x_j}\partial^{s-t}_{p_j}B)\,.
\end{align}

In this context, the Wigner-Weyl transformation is used to systematically recover the classical limit. 
Effectively, it reduces to replacing commutators with Moyal brackets $\frac{1}{i\hbar}[\hat{H},\cdot] \rightarrow \moyb{H}{\cdot}$ and density matrix $\hat{\rho}$ with the Wigner function $W$. The Moyal bracket has the following form in terms of the \emph{star product}
\begin{align}
    \moyb{A}{B} &= \frac{1}{i \hbar}(A \star B - B \star A)\,,
     \label{star product}
\end{align}
Using the $\hbar$-series expansion, the Moyal bracket admits the compact expression
\begin{align}
    \moyb{A}{B} = \frac{2}{\hbar} A \sin\left(\frac{\hbar}{2}\Lambda\right) B\,,
\end{align}
where the Poisson operator reads $\Lambda = \sum_{j = 1}^{N}\left(\overleftarrow{\partial}_{x_j}\overrightarrow{\partial}_{p_j} - \overleftarrow{\partial}_{p_j}\overrightarrow{\partial}_{x_j}\right)$. 
This approach allows one to represent master equations in phase space, as often done to study quantum Brownian motion; see,  e.g.,  \cite{Agarwal69,Dekker77,Caldeira83,Unruh89, HuPazZhang92,Isar96,Habib:1998ai,Cabrera15,Bondar2016,Braasch19,Brody25,Steuernagel15}. 

The GKSL equation in phase space reads 
\begin{eqnarray}
 \partial_t W &=& \{\{H,W\}\}  +
 \sum_{m}\Gamma_m\bigg[L_m\star W\star L_m^*\\
 & & -\frac{1}{2}\left(L_m^*\star L_m\star W+W\star L_m^*\star L_m\right)\bigg]\,,\nonumber
\end{eqnarray}
where $A^*$ denotes the complex conjugate of $A$.

To leading order in the $\hbar$-expansion, the Moyal bracket reduces to \begin{eqnarray}
\moyb{A}{B} = A\Lambda B+ \mathcal{O}(\hbar^2)= \{A,B\}_{\rm P} + \mathcal{O}(\hbar^2),    
\end{eqnarray}
where $A\Lambda B \equiv \{A,B\}_{\rm P}$ is the usual Poisson bracket. In addition, 
\begin{eqnarray}
A\star B=AB+\mathcal{O}(\hbar)\,.\label{clstareq}
\end{eqnarray}
In this limit, the phase-space dissipator identically vanishes if $\Gamma_m$ and $\hat{L}_m$ do not scale with $\hbar$. This is the case as $L_m\star W\star L_m^*\mapsto |L_m|^2W$ and $\frac{1}{2}\left(L_m^*\star L_m\star W+W\star L_m^*\star L_m\right)\mapsto  |L_m|^2W$.

As a specific case, for Hermitian operators $\hat{L}_m$, one finds that the dephasing master equation (\ref{DMEQ}) takes the phase-space form
\begin{eqnarray}
& &  \partial_t W = \{\{H,W\}\}  +\frac{\hbar^2}{2}
 \sum_{m}\Gamma_m\{\{L_m,\{\{L_m,W\}\}\}\}\,,
 \nonumber\\
\end{eqnarray}
and the classical limit of the dissipator still vanishes identically.

However, in the case of fluctuating Hamiltonians, $\hat{L}_m$ is an energy operator and $\Gamma_m$ is given by (\ref{Gmeq}). 
The classical limit of dephasing associated with the term of order $\hbar^0$ involves a nonvanishing dissipator
\begin{eqnarray}
& &  \partial_t W = \{H,W\}_{\rm P}  +
 \sum_{m}\frac{\lambda_m}{2}\{L_m,\{L_m,W\}_{\rm P}\}_{\rm P}\,.
\end{eqnarray}
This is consistent with the use of fluctuating Hamiltonians in both the quantum and classical domains.
In particular, consider the master equation (\ref{ME4ED}) to the dynamical equation for the Wigner function under energy dephasing,
\begin{align}
    \partial_t W &= \frac{2}{\hbar}H \sin \left(\frac{\hbar}{2}\Lambda\right)W\notag\\ &+ 2 \Gamma H\sin \left(\frac{\hbar}{2}\Lambda\right)\left[H \sin \left(\frac{\hbar}{2}\Lambda\right) W\right]\,.
\end{align}
 In this limit, we find that the dynamical equation associated with energy dephasing becomes
\begin{align}
    \partial_t W &= \{H,W\}_{\rm P} + \frac{\lambda}{2}\{H,\{H,W\}_{\rm P}\}_{\rm P}\notag\\ &-\frac{\hbar^2}{24}\left[H\Lambda^3 W + \frac{\lambda}{2} \left(H \Lambda (H \Lambda^3 W) + H \Lambda^3 (H \Lambda W)\right)\right]\notag\\ &+ \mathcal{O}(\hbar^4)\,. \label{evol W}
\end{align}
Recognizing the classical Liouvillian associated with Hamiltonian dynamics $\mathcal{L}\ast \equiv \{H,\ast\}_{\rm P}$ and  defining $\gamma = \frac{\lambda}{2}$, the dynamical equation in the classical limit becomes
\begin{align}
    \partial_t W = \mathcal{L}W + \gamma \mathcal{L}^2 W\,. \label{double comm evolution equation}
\end{align}
Since $\mathcal{L}^2$ commutes with $\mathcal{L}$, we can write the full solution as 
\begin{align}
    W(t) = e^{\gamma t \mathcal{L}^2}e^{t \mathcal{L}}W(0)\,.
\end{align}
This expression makes clear that the classical limit of energy dephasing is a deformation of the classical Hamiltonian dynamics associated with the deformation of the generator $\mathcal{L}\rightarrow \mathcal{L}+\gamma \mathcal{L}^2$. 
This can now be applied to specific systems.

\subsection{Heat Kernel Solution and Ehrenfest Equations}
We begin by considering a simple system of a massive particle moving under a potential $V(\hat{x})$. The Hamiltonian and the propagator $\mathcal{L}$ are given by
\begin{align}
    \hat{H} &= \frac{\hat{p}^2}{2 m} + V(\hat{x})\,,\\
    \mathcal{L} &= -\frac{p}{m}\partial_x + V'(x)\partial_p\,.
\end{align}
As shown in Appendix \ref{Appendix Heat Kernel}, the energy-dephasing dynamics governed by $\mathcal{L}+\gamma \mathcal{L}^2$ can be obtained via a heat kernel in terms of Hamiltonian flow,
\begin{align}
    W(x,p,t) = \frac{1}{\sqrt{2\pi \gamma t}}\int_{-\infty}^{\infty}\,
e^{-\frac{(t - u)^2}{2 \gamma t}}
    W(\phi_u(x,p))\mathrm{d}u\label{eq:smear-wigner}\,.
\end{align}
Here, $W(\phi_u(x,p)) = W(x(u),p(u),u)$, with $x(u),p(u)$ corresponding to the solutions of Hamilton's equation with $H$.

Without referring to the explicit form of the Wigner function or the Hamiltonian flow, we can derive the evolution equation for different moments of the canonical observables $x$ and $p$ in phase space. The average values of the observables defined in Eq. \eqref{avg A} satisfy the Ehrenfest equations
\begin{align}
    \frac{d}{dt}\langle x \rangle &= \frac{\langle p \rangle}{m} - \frac{\gamma}{m}\langle V'(x)\rangle\,,\; \\
    \frac{d}{dt} \langle p \rangle &= - \langle V'(x) \rangle - \frac{\gamma}{m} \langle p V''(x) \rangle\;,
\end{align}
which can be derived using the evolution equation for the Wigner function in Eq. \eqref{double comm evolution equation} and integration by parts.

\sect{Simple Harmonic Oscillator} As a concrete working example, we first consider the simple harmonic oscillator. The potential function is given by $V(\hat{x}) = \frac{1}{2}m \omega^2 \hat{x}^2$ corresponding to the frequency $\omega$. The Hamiltonian flow $\Phi_{u}$ of this model is described by the equations
\begin{align}
    \begin{pmatrix}
        x(u) \\
        p(u) 
    \end{pmatrix} = \begin{pmatrix}
        \cos(\omega u) && \frac{1}{m\omega}\sin(\omega u) \\
        -m\omega \sin(\omega u) && \cos(\omega u)
    \end{pmatrix}\begin{pmatrix}
        x(0) \\
        p(0)
    \end{pmatrix}\,.
\end{align}
Naturally, the Wigner function is then given by $W(x(u),p(u))$. This also depends on the initial configuration of the function, for which we choose a Gaussian function with variances $\sigma^2_x$ and $\sigma^2_p$ along the $x$ and $p$ directions, respectively. This yields
\begin{align}
    W(x(u),p(u)) = \mathcal{A}\exp\left[-\frac{(x-x(u))^2}{2 \sigma^2_x} -\frac{(p-p(u))^2}{2 \sigma^2_p} \right]\,,
\end{align}
where $\mathcal{A} = \frac{1}{2 \pi \sigma_x \sigma_p}$ is the normalization constant. The smeared Wigner function, resulting from the convolution with the kernel in \eqref{eq:smear-wigner} takes the form
\begin{align}
    W(x,p,t) &= \frac{\mathcal{A}}{\sqrt{4\pi \gamma t}}\int_{-\infty}^{\infty}\exp\left[-\frac{(t - u)^2}{2 \gamma t}\right]\notag\\&\times\exp\left[-\frac{(x-x(u))^2}{2 \sigma^2_x}-\frac{(p-p(u))^2}{2 \sigma^2_p} \right]\mathrm{d}u\,.\label{eq:smear-wigner-2}
\end{align}
In general, this integral is difficult to compute in closed form. However, it is relatively straightforward to compute the moments of $W(x,p)$. Let us consider the first and second moments $\langle x \rangle, \langle p \rangle$ and $\langle x^2 \rangle, \langle p^2 \rangle, \langle x p \rangle$.  In particular,
\begin{align}
    \langle x \rangle = e^{-\gamma \omega^2 t}\left[x(0) \cos(\omega t) + \frac{p(0)}{m\omega} \sin{(\omega t)}\right]
\end{align}
and 
\begin{align}
    \frac{\langle p \rangle}{m\omega} = e^{-\gamma \omega^2 t}\left[-x(0) \cos(\omega t) + \frac{p(0)}{m\omega} \sin{(\omega t)}\right]\,.
\end{align}
Note that in the limit $t \rightarrow \infty$, the Wigner function is centered at $\langle x \rangle = 0$ and $\langle p \rangle = 0$ for finite $\gamma$. Next, we compute the second moments, which are the elements of the covariance matrix. One finds 
\begin{align}
    &\langle x^2 \rangle 
    = e^{-2t\gamma\omega^2}\left[\left(\frac{x(0)^2}{2}-\frac{p(0)^2}{2 m^2\omega^2}\right)\cos(2\omega t)\right] + \sigma_x^2\notag\\
    &+ \frac{x(0)p(0)}{2 m \omega}e^{-2 t\gamma\omega^2}\sin(2\omega t) + \frac{x(0)^2}{2} + \frac{p(0)^2}{2m^2\omega^2}\,.
\end{align}
Similarly, for the momentum $p$ we obtain the moment $\langle p^2 \rangle/(m^2\omega^2) = - \langle x^2 \rangle + \sigma^2_x + \sigma^2_p$. As for the cross term, it is given by
\begin{align}
    \frac{\langle x p \rangle}{m\omega} 
    &= \left(\frac{p^2(0)}{m^2 \omega^2} - x(0)^2\right)e^{-2t \gamma \omega^2}\sin(2\omega t)\notag\\
    &+ \frac{x(0)p(0)}{m\omega}\left(1 - 2 e^{-2 t \gamma \omega^2}\cos(2\omega t)\right)\,. 
\end{align}
Note that the covariance matrix tends to a constant independent of $t$ and $\gamma$ in the limit $t\rightarrow \infty$ for any finite $\gamma$. Thus, the final Wigner function at late times, representing the steady state, has a finite spread around the center $(x, p) =(0, 0)$, irrespective of the starting non-zero $(x(0),p(0))$.

\section{Double-Anticommutator Master Equation and Its Classical Limit \label{SecDAME}}

Next, we investigate the evolution encoded in a quantum master equation involving a double anticommutator, recently derived \cite{pablo2025quantum} in the context of fluctuating non-Hermitian systems \cite{Pearle89,Wang25a,Wang25b}. 
Specifically, we consider a double anticommutator with the Hamiltonian, supplemented by a trace-restoring term inspired by ``balanced gain and loss" models. Such evolution is not of the GKSL form and is nonlinear in the quantum state. The quantum master equation reads
\begin{align}
    \frac{d\hat{\rho}}{d t} = -\frac{i}{ \hbar}[\hat{H},\hat{\rho}] - \Gamma\{\hat{H},\{\hat{H},\hat{\rho}\}\} + 4 \Gamma \Tr(\hat{H}^2 \hat{\rho})\hat{\rho}\,,  \label{DC Master eqn}
\end{align}
where $\{\hat{A},\hat{B}\}= \hat{A}\hat{B} + \hat{B}\hat{A}$ is the anticommutator. The last term, which is nonlinear in $\hat{\rho}$, ensures that the trace of $\hat{\rho}$ is preserved over time.
Our goal is to understand the classical limit of this dynamics.

We start by considering its phase-space representation, making use of the symmetric Moyal bracket $\moybst{A}{B} = A \star B + B \star A$, 
\begin{equation}
    \frac{d W}{d t} = \moyb{H}{W} - \Gamma\moybst{H}{\moybst{H}{W}} + 4\Gamma \langle H \star H\rangle W\,,
\end{equation}
where $\Tr(\hat{H}^2 \hat{\rho}) = \langle H\star H\rangle$. 
In what follows, it is insightful to keep the expansion up to $\mathcal{O}(\hbar^4)$. We have already considered the $\hbar$-expansion of the first term $\moyb{H}{W}=\{H,W\}_{\rm P}-\frac{\hbar^2}{24}H\Lambda^3 W+\mathcal{O}(\hbar^4)$. As for the symmetric Moyal bracket,  
\begin{align}
    \moybst{H}{W} &= 2 H \cos\left(\frac{\hbar}{2}\Lambda\right) W \notag\\ 
    &= 2 H W - \frac{\hbar^2}{4}H \Lambda^2 W + \mathcal{O}(\hbar^4)\,, \label{symm Moyal Bracket}
\end{align}
whence it follows that
\begin{align}
    &\moybst{H}{\moybst{H}{W}}\notag\\ %
    &= 4 H^2 W -\frac{\hbar^2}{2}[H \Lambda^2 H W  + H (H \Lambda^2 W)] + \mathcal{O}(\hbar^4)\,. 
\end{align}
Finally, the expectation value $\langle H \star H \rangle$ simply yields
\begin{align}
    \langle H \star H \rangle = \langle H^2 \rangle + \frac{i \hbar}{2} \langle H \Lambda H\rangle - \frac{\hbar^2}{8}\langle H \Lambda^2 H \rangle + \mathcal{O}(\hbar^4)\,.
\end{align}
As the linear term in $\hbar$ vanishes, 
\begin{align}
    \langle H \star H \rangle =  \langle H^2 \rangle  - \frac{\hbar^2}{8}\langle H \Lambda^2 H\rangle + \mathcal{O}(\hbar^4)\,.
\end{align}
Using this, the effective classical master equation can be written as 
\begin{align}
    \frac{d W}{d t} &= \{H,W\}_{\rm P} - 4\Gamma\left(H^2 - \langle H^2 \rangle\right)W \notag\\ 
    &-\frac{\hbar^2}{24}H\Lambda^3 W+ \frac{\Gamma\hbar^2}{2} \left[H \Lambda^2 H W + H(H \Lambda^2 W)\right]\notag\\
    &- \frac{\hbar^2 \Gamma}{2}\langle H \Lambda^2 H \rangle W + \mathcal{O}(\hbar^4)\,. \label{W evol double AC}
\end{align}
The terms proportional to $\hbar^0$ are $\{H,W\}_{\rm P} - 4 \Gamma (H^2 - \langle H \rangle^2)W$, which represent the Poisson-bracket propagation altered by the balanced gain/loss contribution proportional to $\Gamma$. Since the highest order terms in this expansion contain $\Lambda^2$ and $\Lambda^3$, it is worth writing down the corresponding explicit expressions 
\begin{align}
    H \Lambda^2 W &= \partial^2_x H \partial^2_p W + \partial^2_p H \partial^2_x W\notag\\ &- 2 \left(\partial_x \partial_p H\right) \left(\partial_x \partial_p W\right)\,,\\ 
    H \Lambda^3 W &= \partial^3_x H \partial^3_p W - \partial^3_p H \partial^3_x W\notag\\ &- 3 \left(\partial_x^2 \partial_p H\right) \left(\partial_x \partial_p^2 W\right)+3 \left(\partial_x \partial_p^2 H\right) \left(\partial_x^2 \partial_p W\right)\,,\\
    H \Lambda^2 H W &= \partial^2_x H \partial^2_p (HW) + \partial^2_p H \partial^2_x (HW) \notag\\ &- 2 \left(\partial_x \partial_p H\right) \left(\partial_x \partial_p HW\right)\,.
\end{align}
We clarify that the above analysis relies on  $\Gamma$ being independent of $\hbar$. By contrast, the scaling in \eqref{Gmeq} would rescale the terms in the  $\hbar$-expansion by $1/\hbar^2$, causing the nonlinear term to diverge in the classical limit. Such a divergence leads to constrained dynamics on a surface in phase space with constant $H(\mathbf{x},\mathbf{p})^2$, i.e., on the energy shell. 

\subsection{Purely Classical Limit}
Let us consider the $\hbar^0$ case of this equation. This is given by
\begin{align}
    \frac{d W}{d t} = \{H, W\}_{\rm P} - 4 \Gamma (H^2 - \langle H^2 \rangle)W\,, \label{cl_eq_DAC}
\end{align}
where $\langle H^2 \rangle = \int H^2(x,p) W(x,p,t) \mathrm{d}x\mathrm{d}p$. This equation preserves the normalization as $\frac{d}{d t}\int W \mathrm{d}x\mathrm{d}p = 0$. In general, such equations are difficult to solve. However, it is possible to construct solutions along characteristic curves $(x(t), p(t))$ which satisfy Hamilton's equations. Along these curves, the Hamiltonian is conserved, i.e., $H(x(t),p(t)) = H(x(0),p(0))$. Let us denote the coordinates along the characteristic curves as $\phi_t := (x(t),p(t))$. Thus, $H(\phi_t) = H(\phi_0) \equiv h$, assuming that $H$ does not have an explicit time dependence. Therefore, the equation for $W$ reduces to 
\begin{align}
    \frac{d}{dt}W(\phi_t,t) = - 4 \Gamma (h^2 - \langle H^2 (t) \rangle)W(\phi_t,t)\,.
\end{align}
This equation can be formally integrated to give the following solution
\begin{align}
    W(\phi_t,t) =  W(\phi_0,0)e^{-4 \Gamma h^2 t}\exp \left(4 \Gamma \int_0^t\langle H^2 (s) \rangle \mathrm{d}s \right)\,. \label{Wig Solution}
\end{align}

Note that the integral over the phase space is equivalent to the integral over all initial conditions of the characteristic curve $\phi_0$. Using this, we can compute the average $\langle H^2 (t)\rangle$. This involves plugging in the formal solution 
\begin{align}
    \langle H^2 (t)\rangle &= \int h^2 W(\phi_t,t)\mathrm{d}\phi_0 \notag\\
    &= \int h^2 W(\phi_0,0)e^{-4 \Gamma h^2 t}e^{4 \Gamma \int_0^t\langle H^2 (s) \rangle \mathrm{d}s}\mathrm{d}\phi_0 \notag\\
    &= e^{4 \Gamma \int_0^t \langle H^2 (s)\rangle\mathrm{d}s}\int h^2 W(\phi_0,0)e^{-4 \Gamma h^2 t}\mathrm{d}\phi_0 \notag\\
    &\equiv A(t)M(t)\,,
\end{align}
where we have introduced the notation $A(t) = e^{4 \Gamma \int_0^t \langle H^2 (s)\rangle\mathrm{d}s}$ and $M(t) = \int h^2 W(\phi_0,0)e^{-4 \Gamma h^2 t}\mathrm{d}\phi_0$. The equation for $A(t)$ can then be written as
\begin{align}
    \frac{d}{dt}A(t) = 4\Gamma M(t) A^2(t)\,,
\end{align}
which has the solution $1 - A^{-1}(t) = 4\Gamma\int_0^t M(s)\mathrm{d}s$, where we have used the fact that $A(0) = 1$. Therefore, we can write
\begin{align}
    A(t) = \frac{1}{1 - 4 \Gamma \int_0^t M(s)\mathrm{d}s}\,,
\end{align}
and the full solution of the Wigner function can be written as
\begin{align}
    W(\phi_t,t) = W(\phi_0,0)A(t)e^{-4\Gamma h ^2 t}\,.
\end{align}
The time integral in the denominator can be evaluated explicitly,  
\begin{align}
    \int_0^t M(s)\mathrm{d}s = \int h^2 W(\phi_0,0)\frac{1 - e^{-4 \Gamma h^2 t}}{4 h^2 \Gamma}\mathrm{d}\phi_0\,.
\end{align}
This leads us to the simplified expression
\begin{align}
    W(\phi_t,t) = \frac{W(\phi_0,0)e^{-4\Gamma h^2 t}}{\int W(\phi_0,0)e^{-4\Gamma h^2 t}\mathrm{d}\phi_0}\,.
\end{align}
Note that this solution is along the characteristic curve. Therefore, the actual solution is obtained by reversing the transformation to the characteristic curve, which gives us the following relation
\begin{align}
    W(x,p,t) = \frac{W(\phi_{-t})e^{-4 \Gamma H^2(x,p) t}}{\int W(\phi_{-t})e^{-4 \Gamma H^2(x,p) t}\mathrm{d}x\mathrm{d}p}\,,
\end{align}
where $W(\phi_t)$ is the term generated by the Hamiltonian flow. In the limit $\Gamma \rightarrow 0$, it reduces to the usual Hamiltonian solution form. 

\sect{Energy moments} We can also compute the $n-$th order energy moments $\mu_n$ along the characteristic curves, defined as
\begin{align}
    \mu_n = \int h^n W(\phi_t, t)\mathrm{d}x\mathrm{d}p\,.
\end{align}
The rate of change with respect to time satisfies a recursion relation
\begin{align}
    \frac{d}{dt} \mu_n &= - 4 \Gamma\int h^n  (h^2 - \langle H^2 (t) \rangle)W(\phi_t,t) \mathrm{d}x\mathrm{d}p\nonumber\\ 
    &= -4 \Gamma [\mu_{n+2}(t) - \mu_n(t)\mu_2(t)]\,.
\end{align}
We first consider the case $n=0$, associated with the normalization, to understand how the total area under the Wigner function evolves. It satisfies the differential equation
\begin{align}
    \frac{d}{dt}\mu_0 (t)= 4\Gamma \mu_2(t) (\mu_0(t)-1)\,,
\end{align}
which can be solved to get
\begin{align}
    \mu_0(t) = 1 + (\mu_0(0)-1)\exp\left(4\gamma \int_0^t \mu_2(s)\mathrm{d}s\right)\,.
\end{align}
The total area under the Wigner function is conserved if the initial state is normalized. If not, there is an exponential drift governed by the second moment. The second moment itself satisfies the differential equation
\begin{align}
    \frac{d}{dt}\mu_2(t) = -4 \Gamma[\mu_4(t) - (\mu_2(t))^2]\,.
\end{align}
Using the Cauchy-Schwarz inequality, we see that $\mu_4(t) \geq (\mu_2(t))^2$. Indeed, the rate is given by the variance $\text{Var}(h^2)=\mu_4(t) - (\mu_2(t))^2 \geq 0$, which implies that the second moment decreases monotonically.  The evolution drives the Wigner function to minimize $\mu_2$, as is also evident from Eq. \eqref{cl_eq_DAC}, which enhances regions where $H^2 < \langle H^2 \rangle$. Before discussing specific examples, we next show that the dynamics admits a gradient flow description.

\section{Gradient Flow Representation of Double Bracket Equations} \label{SecGradfow}

\begin{figure}[h]
    \centering
    \begin{tikzpicture}[scale=0.65]
        
        \path[] (1,1) to[out=-10,in=150] (5,-1);
        \path[name path=border1] (0,0) to[out=-10,in=150] (6,-2);

        \path[name path=border2] (12,1) to[out=150,in=-10] (5.5,3.2);
        \path[name path=redline] (0,-0.4) -- (12,1.5);
        \path[name intersections={of=border1 and redline,by={a}}];
        \path[name intersections={of=border2 and redline,by={b}}];

        \shade[left color=blue!50!black!40,right color=green!50!black!40] 
          (0,0) to[out=-10,in=150] (6,-2) -- (12,1) to[out=150,in=-10] (5.5,3.7) -- cycle;

        \draw[red!50!black!100,line width=1.5pt,shorten >= 3pt,shorten <= 3pt] 
          (a) .. controls (6,-0.2) and (5,3.5) ..
          coordinate[pos=0.22] (cux1) 
          coordinate[pos=0.57] (cux2) 
          coordinate[pos=0.88] (cux3) (b);
        
        \begin{scope}[
        every node/.style={
          inner sep=0pt,
          opacity=0.4,
          single arrow,
          draw=none,
          single arrow head extend=2pt}
        ];

        \begin{scope}[tdplot_main_coords]
            \foreach \x in {0,1,...,25} {
        \draw[black, thin, opacity=0.4] (\x/5+2,5,0) -- (\x/5+2,10,0);
        }
        \foreach \y in {0,1,...,25} {
            \draw[black, thin, opacity=0.4] (2,\y/5+5,0) -- (7,\y/5+5,0);
        }
        
        \shade [shading=radial, 
                inner color=white, 
                outer color=red!30!white, 
                opacity=0.8] (4.2,7) circle [radius=0.1cm];

        \shade [shading=radial, 
                inner color=white, 
                outer color=black!30!white, 
                opacity=0.8] (2.2,7) circle [radius=0.1cm];
        
        \end{scope}

        
        \end{scope}
        
        \node[black, thick]  at (5.0,2.9) {$[\hat{X},\hat{\sigma}_t]$};
        \node[black, thick]  at (6,-0.7) {\Large$\mathcal{M}$};
        \node[black,thick] at (10.2,3.3) {\Large$T_{\hat{\sigma}_t}$};
        \node[red!50!black!100, thick]  at (4.2,1.0) {\Large$\hat{\mathcal{O}}$};
        \node[red!50!black!100, thick]  at (7.2,1.6) {\Large$\hat{\sigma}_t$};
    \end{tikzpicture}
    \caption{Gradient flow description of the time evolution, showing the orbit associated with the trajectory $\hat{\mathcal{O}}$ in the manifold $\mathcal{M}$. The tangent space,  shown as the plane surface with grids, is denoted by $T_{\hat{\sigma}_t}$, at the point $\hat{\sigma}_t$ on $\hat{\mathcal{O}}$, for some parameter $t$ parametrizing the trajectory. An element on $T_{\hat{\sigma}_t}$ is marked as $[\hat{X},\hat{\sigma}_t]$.}
    \label{fig:schematic}
\end{figure}
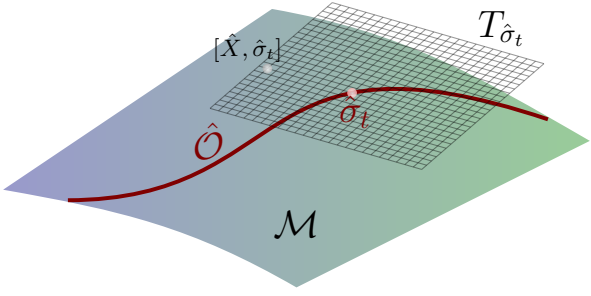

The dynamics of a quantum state under commutator or double-commutator dissipators can be described as a gradient flow on some constrained manifold. This approach~\cite{rossmann2023,Helmke_Moore_2014,Bloch_1994,helmkereview} offers several advantages for advanced methods in quantum control, optimization and machine learning~\cite{Wiersma2023optimizing, Wiersema2024here,xiaoyue2024strategiesoptimizingdoublebracketquantum,Gluza2024doublebracket,mcmahon2025equatingquantumimaginarytime,villanueva2025}. The canonical formulation of gradient flow is targeted towards dynamics of the form $\partial_t \rho_t = [H,\rho_t]$ or $\partial_t\rho_t = [[H,\rho_t],\rho_t]$ (or combinations thereof), which are useful for diagonalization, for example. The main idea is the following: consider a surface $\mathcal{M}$ where each point denotes a density matrix $\hat{\sigma}$ allowed in the Hilbert space of the problem. Under some generator of dynamics $\mathcal{L}$, a given density matrix $\hat{\sigma}_0$ maps a trajectory on this surface. This trajectory is denoted by $\hat{\mathcal{O}}$. The collection of all such trajectories is equal to $\mathcal{M}$. The tangent space $T_{\hat{\sigma}_t}$ at any point $\hat{\sigma}_t$ along the trajectory is the space of all commutators $[\hat{X},\hat{\sigma}_t]$ for all $n$-dimensional matrices $\hat{X} \in \mathbb{R}^{n \times n}$; see Fig. \ref{fig:schematic}. 

It was demonstrated in~\cite{Bloch1997} that double-bracket generators $\mathcal{L}\rho_t = [[H,\rho_t],\rho_t]$ can be written as a gradient of a scalar function (potential) $\mathbf{\Phi}(\hat{\sigma})$ that is defined on the tangent manifold through an appropriate Riemannian metric, where the gradient is evaluated at $\hat{\sigma} = \hat{\rho}_t$. In other words, for the isospectral double bracket, we can write
\begin{align}
    \frac{d\hat{\rho}_t}{dt} = [[H,\hat{\rho}_t],\hat{\rho}_t] = -\nabla_{\hat{\sigma}} \mathbf{\Phi}(\hat{\sigma})\Big\vert_{\hat{\sigma} = \hat{\rho}_t}\,.
\end{align}
In the case of such isospectral double bracket flows, it is known that the potential is given by 
\begin{align}
    \mathbf{\Phi}(\hat{\sigma}) = \frac{1}{2}\vert\vert \hat{H} - \hat{\sigma}\vert\vert^2_{\text{HS}}\,,
\end{align}
where $\vert\vert \hat{A} \vert\vert_{\rm HS}=\sqrt{\Tr \hat{A}^\dag \hat{A}}$ stands for the Hilbert-Schmidt norm. In performing this computation, note that the matrix calculus involves  $d_{\hat{\sigma}} \mathbf{\Phi} = \Tr[\nabla_{\hat{\sigma}}\mathbf{\Phi}(\hat{\rho},\hat{\sigma}) d\hat{\sigma}]$. From this expression, we can then read off $\nabla_{\hat{\sigma}}\mathbf{\Phi}(\hat{\rho},\hat{\sigma})$. Similar gradient flows can be written for master equations involving balanced gain and loss ~\cite{Mittnenzweig2017,carlen2017,cao2019gradient,carlen2020}, while for more general Lindbladians, nearly-gradient flows can be derived~\cite{Kaplanek2025Gradient}.

In this work, we are interested in the dynamics of a state $\rho_t$ under single- and double-bracket master equations $\partial_t \rho_t = \mathcal{L}\rho_t$, where the brackets can be commutators or anti-commutators. Furthermore, the generator $\mathcal{L}\rho_t$ can be linear or nonlinear in $\rho_t$, making it different from the isospectral diagonalizing flow. Regardless, the time evolution can still be described as a gradient flow by introducing appropriate scalar potentials. We are interested in equations of the following form
\begin{align}
    \frac{d\hat{\rho}}{dt} = \mathcal{L}\hat{\rho} - \Tr\left(\mathcal{L}\hat{\rho}\right)\hat{\rho}\,,
\end{align}
with a generator $\mathcal{L}$ of the form
\begin{align}
    \mathcal{L}\hat{\rho} = -\frac{i}{\hbar}[\hat{H},\hat{\rho}] &- \sum_m \xi_m \{\hat{K}_m,\{\hat{K}_m,\hat{\rho}\}\}\notag\\ &- \sum_n \Gamma_n [\hat{L}_n,[\hat{L}_n,\hat{\rho}]]\,.
\end{align}
Here, $\hat{L}_n$ and $\hat{K}_m$ are general (Hermitian) operators. 
As we next show, the action of the Linbladian can be recast in terms of the gradient of a (scalar) potential function $\mathbf{\Phi}$, that can be split into three terms
\begin{align}
\mathbf{\Phi}(\hat{\rho},\hat{\sigma}) = \mathbf{\Phi}_0(\hat{\rho},\hat{\sigma}) + \mathbf{\Phi}_{+}(\hat{\sigma}) + \mathbf{\Phi}_{-}(\hat{\sigma})\,. 
\end{align}

The commutator term in $\mathcal{L}\hat{\rho}$ naturally lends itself to a Hamiltonian flow, rather than a gradient flow. However, using the knowledge of the density matrix $\hat{\rho}$, it is possible to construct a function $\mathbf{\Phi}_0(\hat{\rho},\hat{\sigma})$ where the gradient is taken with respect to $\hat{\sigma}$. This is given by 
\begin{align}
    \mathbf{\Phi}_0(\hat{\rho},\hat{\sigma}) = \Tr\left(-\frac{i}{\hbar}[\hat{H},\hat{\rho}]\hat{\sigma}\right)\,.
\end{align}
In terms of the underlying Lie group manifolds, this is expected since the commutator is precisely the tangent vector at the point $\hat{\sigma} = \hat{\rho}$ on the tangent space $T_{\hat{\sigma}} \mathcal{O}_{\hat{\sigma}}$ of the orbit $\mathcal{O}_{\hat{\sigma}}$ of $\hat{\sigma}$. The double commutator admits the known gradient flow representation with the function
\begin{align}
    \mathbf{\Phi}_{-}(\hat{\sigma}) &= -\sum_n \Gamma_n \Tr\left([\hat{L}_n,\hat{\sigma}][\hat{L}_n,\hat{\sigma}]^\dagger\right)\notag\\ &\equiv-\sum_n \Gamma_n \vert\vert [\hat{L}_n,\hat{\sigma}]\vert\vert_{\rm HS}^2\,.
\end{align}
This form is familiar in the study of decoherence,  as it governs the short-time decay of the purity and fidelity under dephasing dynamics \cite{Lidar98,Beau17}. It can be rewritten as a generalized covariance and equals to the decoherence rate, which generalizes the celebrated estimate by Zurek for the case of high-temperature quantum Brownian motion \cite{Zurek91} to generic dephasing evolutions. As such, it has manifold applications, e.g., in noisy quantum systems \cite{chenu2017quantum}, nonexponential decay \cite{Beau17}, quantum metrology \cite{Beau17metro}, and dissipative quantum chaos \cite{xu2019extreme,YangXu24}.

The third term in $\mathcal{L}$, involving the anti-commutator, corresponds to the potential
\begin{align}
    \mathbf{\Phi}_{+}(\hat{\sigma}) &= -\sum_m \xi_m \Tr\left(\{\hat{K}_m,\hat{\sigma}\}\{\hat{K}_m,\hat{\sigma}\}^\dagger\right)\notag\\
    &\equiv-\sum_n \xi_m \vert\vert \{\hat{K}_m,\hat{\sigma}\}\vert\vert_{\rm HS}^2\,.
\end{align}
This demonstrates that the anticommutator term has a gradient interpretation. To interpret this potential in the manifold picture, the anticommutator $\{\hat{K}_m,\hat{\sigma}\}$ has to be interpreted as a commutator $[\hat{K}'_m,\hat{\sigma}]$ in terms of some other operator $\hat{K}'_m$ that allows us to write $\vert\vert \{\hat{K}_m,\hat{\sigma}\}\vert\vert_{\rm HS}^2 = \vert\vert [\hat{K}'_m,\hat{\sigma}]\vert\vert_{\rm HS}^2$. This is because the product structure on Lie group manifolds is defined as $X \circ Y = [X,Y]$ and is essential to define a uniform metric on the tangent space. We do not comment further on this, aside from noting that one potential way to construct $\hat{K}'_m$ from $\hat{K}_m$ would be through the inverse of the symmetric logarithmic derivative~\cite{braunstein1994statistical}. We note that in such construction $\hat{K}'_m$ is in general a function of $\hat{\sigma}$.
 
Retaining the commutator and anti-commutators independently, we can write the dynamics $\mathcal{L}\hat{\rho}$ as the following gradient flow,
\begin{align}
    \mathcal{L}\hat{\rho} = -\nabla_{\hat{\sigma}}\mathbf{\Phi}(\hat{\rho},\hat{\sigma})\vert_{\hat{\sigma} = \hat{\rho}}\,.
\end{align}
The last term that we have to address is the trace-correcting nonlinear term $\Tr(\mathcal{L}\hat{\rho})\hat{\rho}$. As an upshot,
\begin{align}
    \frac{d\hat{\rho}}{dt} = &-\nabla_{\hat{\sigma}}\mathbf{\Phi}(\hat{\rho},\hat{\sigma})\vert_{\hat{\sigma} = \hat{\rho}}\notag\\ &- \frac{1}{2}\Tr\left[\nabla_{\hat{\sigma}}\mathbf{\Phi}(\hat{\rho},\hat{\sigma})\vert_{\hat{\sigma} = \hat{\rho}}\right]\nabla_{\hat{\sigma}}\Tr(\hat{\sigma}^2)\vert_{\hat{\sigma} = \hat{\rho}}\,.
\end{align}
Note that both with and without the trace preserving term, the steady state $\hat{\rho}_{ss}$ is given by the inflection points of the function $\mathbf{\Phi}(\hat{\rho},\hat{\sigma})$. The trace-preserving case can be understood as a type of a projected gradient flow, where the operation $P_{\hat{\sigma}}(\hat{A})$ projects the operator $\hat{A}$ onto the tangent space of all operators with unit trace. In terms of the projection, we can write the dynamics as
\begin{align}
    \frac{d\hat{\rho}}{dt} = -P_{\hat{\sigma}}\left[ \nabla_{\hat{\sigma}}\mathbf{\Phi}(\hat{\rho},\hat{\sigma})\vert_{\hat{\sigma} = \hat{\rho}}\right]\,.
\end{align}
Therefore, the dynamics can be mapped to a gradient flow with a projective constraint. Let us now focus on the particular case where $n, m = 1$ and $\hat{L} = \hat{H}$ and $\hat{K} = \hat{H}$, which are the cases considered in this work. Assuming the Hermiticity of $\hat{\rho},\hat{\sigma}$, we find that in terms of the eigenbasis $\ket{n}$ of $\hat{H}$, the potential term $\mathbf{\Phi}$ can be written as
\begin{align}
    \mathbf{\Phi}(\hat{\rho},\hat{\sigma}) &= -\frac{i}{\hbar}\sum_{n,m} E_n \left(\rho_{n m}\sigma_{m n} - \sigma_{n m}\rho_{m n}\right) \notag\\
    &- 2\Gamma\sum_{n, m}\left(E^2_n - E_n E_m \right)\sigma_{n m}\sigma_{m n}\notag\\ &- 2\xi\sum_{n m}\left(E^2_n + E_n E_m \right)\sigma_{n m}\sigma_{m n}\,.
\end{align}
The first term is the usual commutator expression, while the second and third terms can be combined into $\mathbf{S}(\hat{\rho},\hat{\sigma})$ where
\begin{align}
    \mathbf{S}(\hat{\rho},\hat{\sigma}) = -\sum_{n,m} 2 E_n \left((\xi + \Gamma)E_n + (\xi - \Gamma)E_m\right)\vert\sigma_{m n}\vert^2\,.
\end{align}
Assuming that $E_n \geq 0\,\forall\, n$, this tells us that the function $\mathbf{S}(\hat{\rho},\hat{\sigma}) \leq 0$ and convex if $\xi > \Gamma$. This ensures the existence of a local minimum and, therefore, a steady state in the absence of the Hamiltonian term. This is a sufficient, but not necessary, condition. 

The semi-classical gradient flow equations are then obtained by replacing the trace $\Tr$ by an integral over the phase space and inserting the Weyl symbol corresponding to the operators within the trace. This yields
\begin{align}
    \mathbf{\Phi}_{-}(W_\sigma) &= -\hbar^2 \sum_n 2\Gamma_n \int \{\{L_n,W_\sigma\}\}^2\mathrm{d}\mathbf{x}\mathrm{d}\mathbf{p},\\
    \mathbf{\Phi}_{+}(W_\sigma) &= \sum_m 2\xi_m \int \{\{K_m,W_\sigma\}\}_{+}^2\mathrm{d}\mathbf{x}\mathrm{d}\mathbf{p}\,. 
\end{align}
These quantities are nonlinear in the Wigner function $W_\sigma$ corresponding to the tunable density matrix $\hat{\sigma}$. The Weyl symbols $L_n, K_n$ are obtained using the definition (\ref{DefWeyl}) with the operators $\hat{L}_n,\hat{K}_n$. The commutator portion $\mathbf{\Phi}_0 (\hat{\rho},\hat{\sigma})$ takes the following form
\begin{align}
    \mathbf{\Phi}_{0}(W_\sigma) = \int \{\{ H,W_\sigma\}\}W_\sigma\mathrm{d}\mathbf{x}\mathrm{d}\mathbf{p}\,.
\end{align}
Combining $\Phi_{0}$ and $\Phi_{\pm}$ gives the Moyal form of the function $\mathbf{\Phi}$. The contribution of these terms can be isolated order-by-order in $\hbar$ from $\mathbf{\Phi}$. These will then correspond to the $\hbar$ expansion of the equation for $\partial_t W$ discussed above.

\section{Examples}\label{SecExamples}

In this section, we look at the effect of the double bracket structure in two different systems. We first consider the simple harmonic oscillator and solve the problem numerically and analytically to characterize the dynamics in the quantum and classical limits. In addition, we use a driven anharmonic oscillator as a toy model to explore the interplay between decoherence and chaos.

\subsection{Simple Harmonic Oscillator}
\begin{figure*}[ht]
    \centering
    \includegraphics[width=0.32\linewidth]{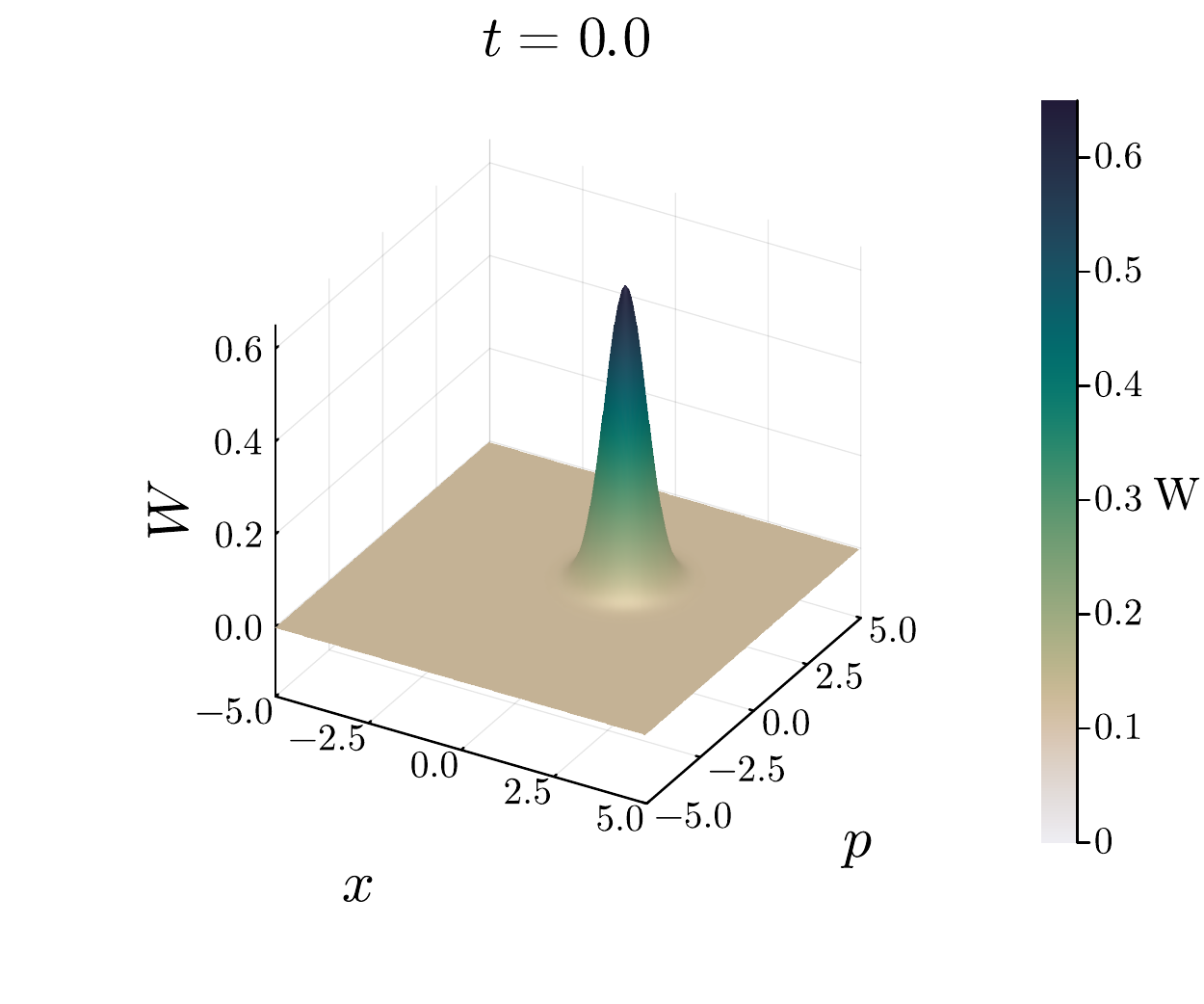}
    \includegraphics[width=0.32\linewidth]
    {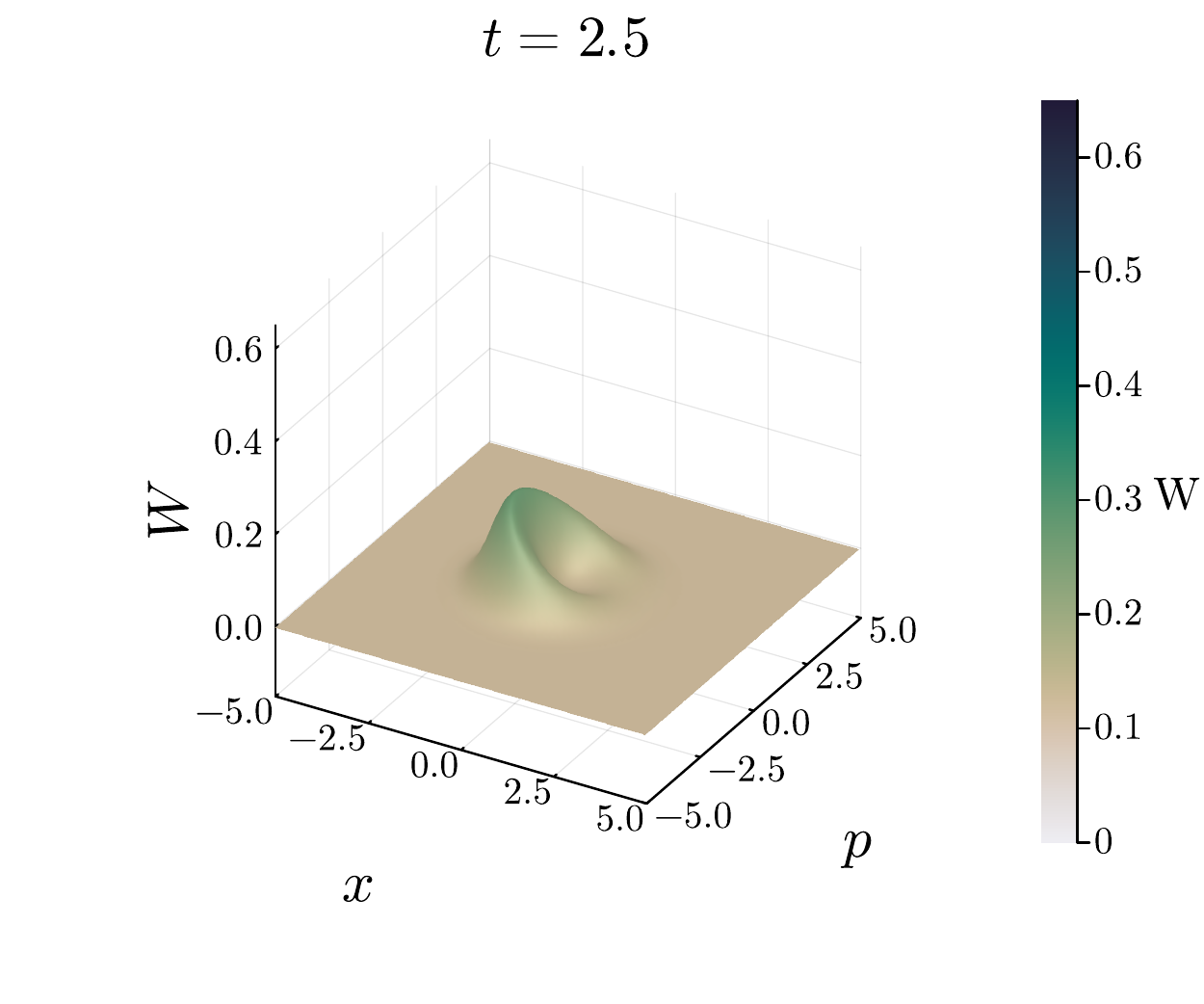}
    \includegraphics[width=0.32\linewidth]{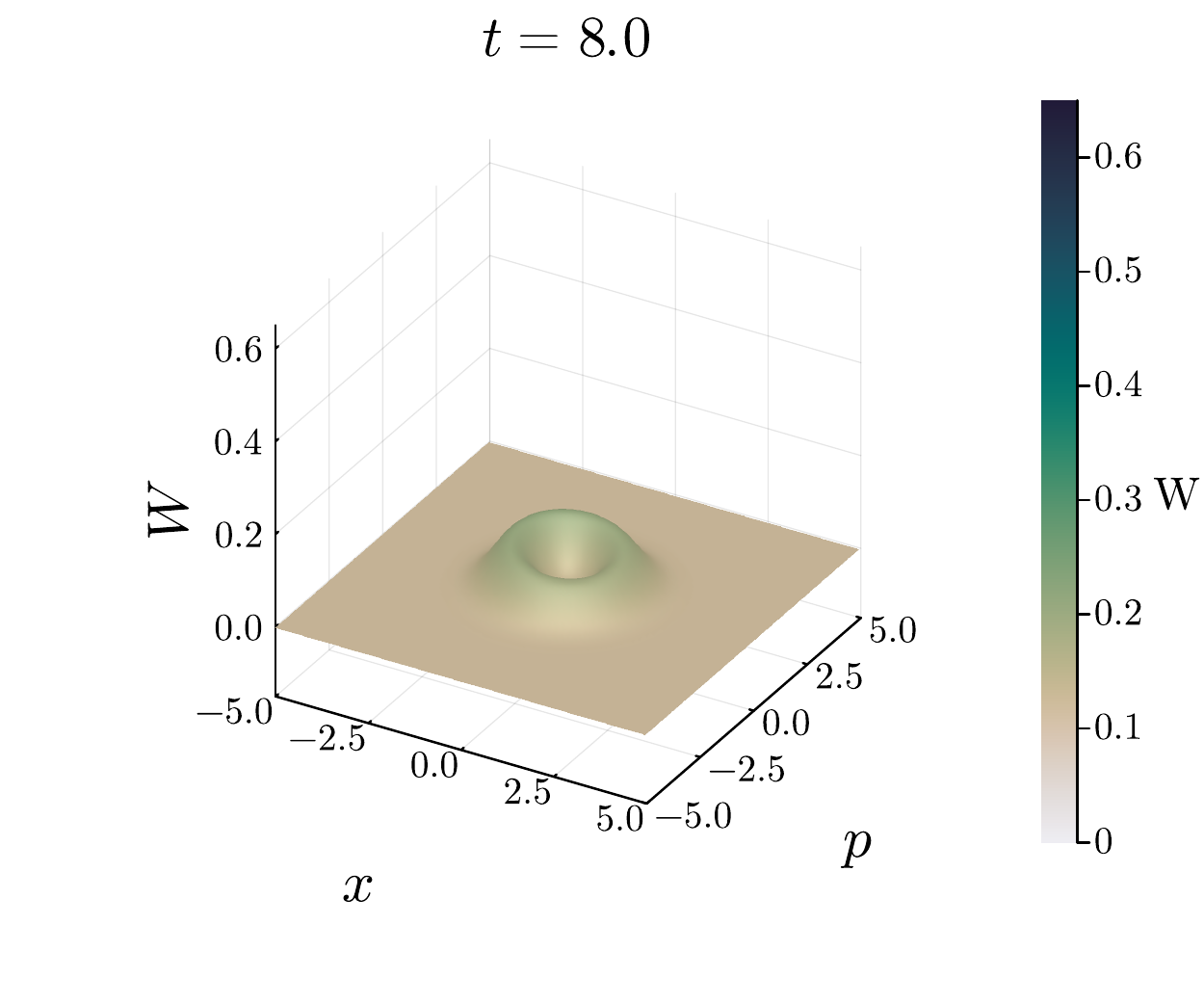}
    \includegraphics[width=0.32\linewidth]{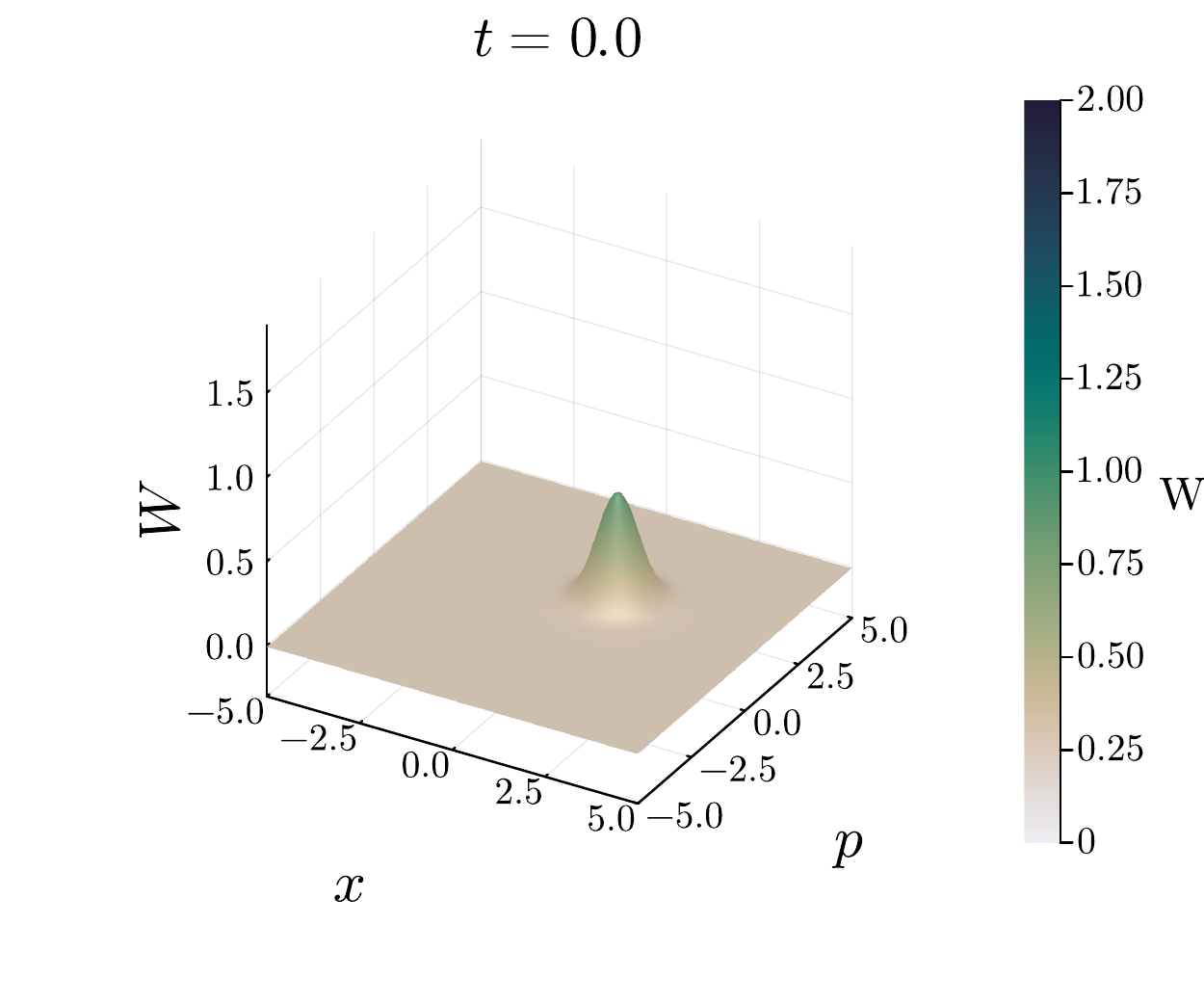}
    \includegraphics[width=0.32\linewidth]{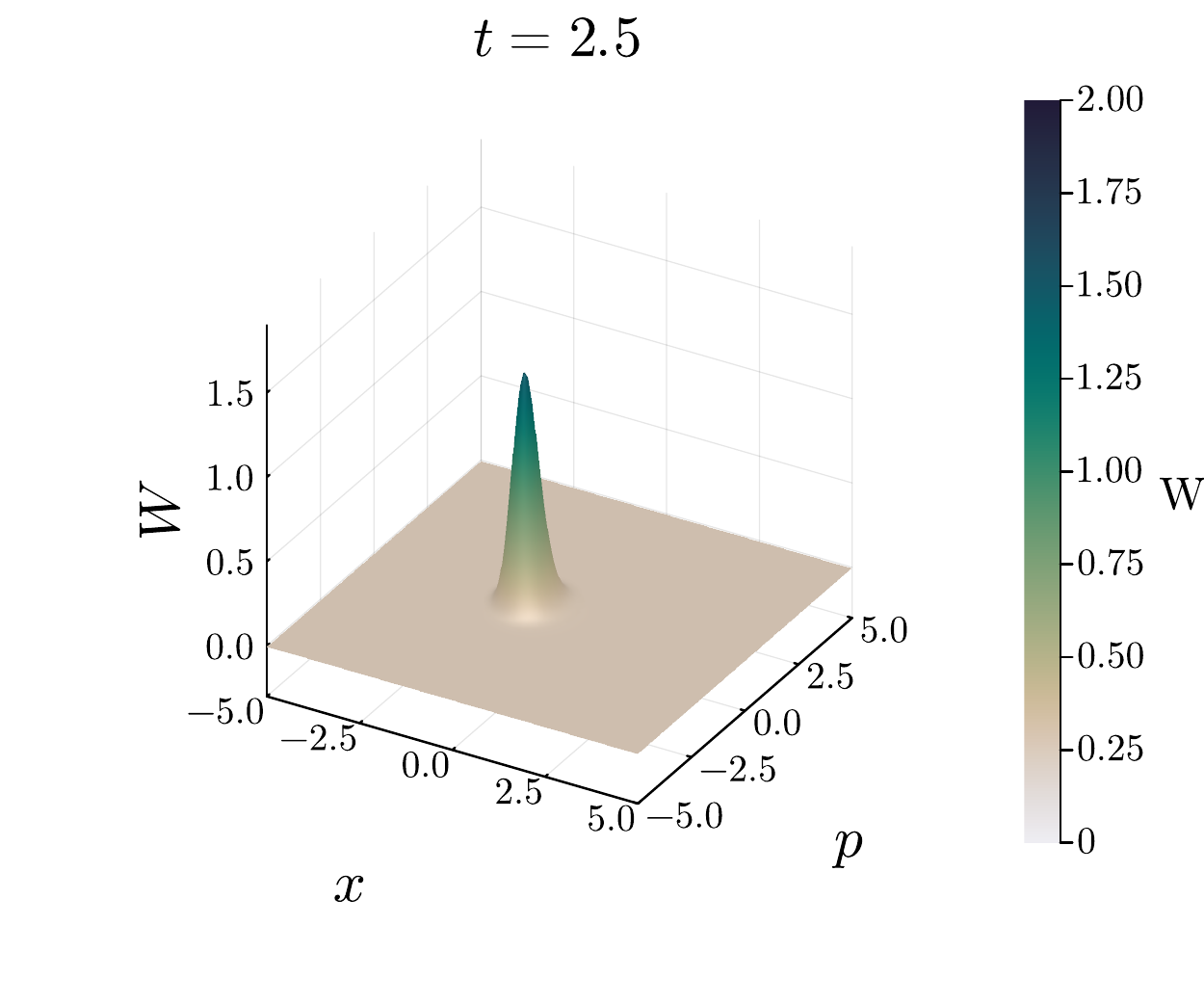}
    \includegraphics[width=0.32\linewidth]{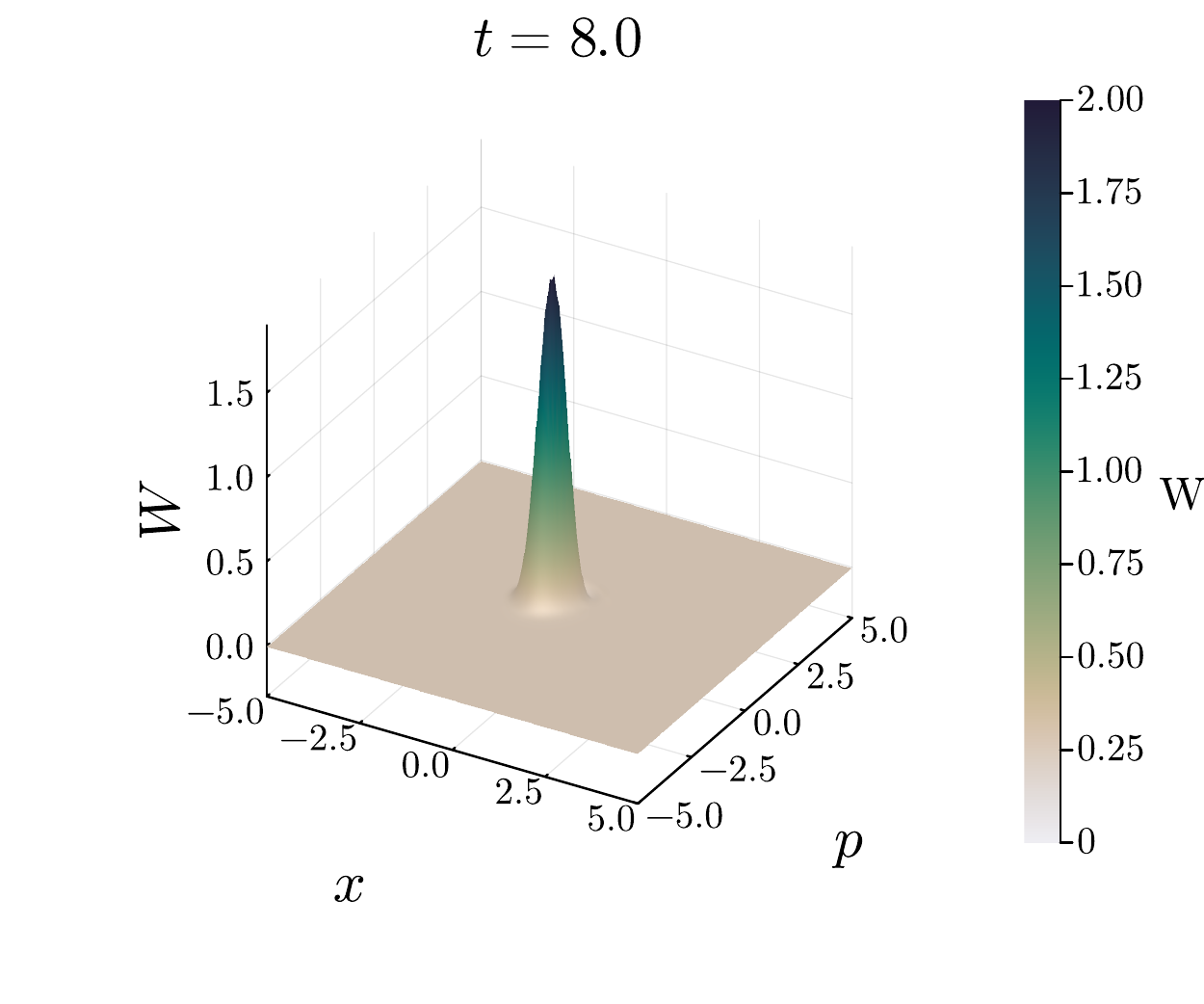}
    \caption{Evolution of the Wigner function of an initial Gaussian wavepacket with mean at $ (x,p) = (1,1)$ in the semiclassical limit of double-commutator (top) and double-anticommutator (bottom) master equations ($\Gamma = 0.3$). }
    \label{fig:1}
\end{figure*}

\begin{figure*}[ht]
    \centering
    \includegraphics[width=0.32\linewidth]{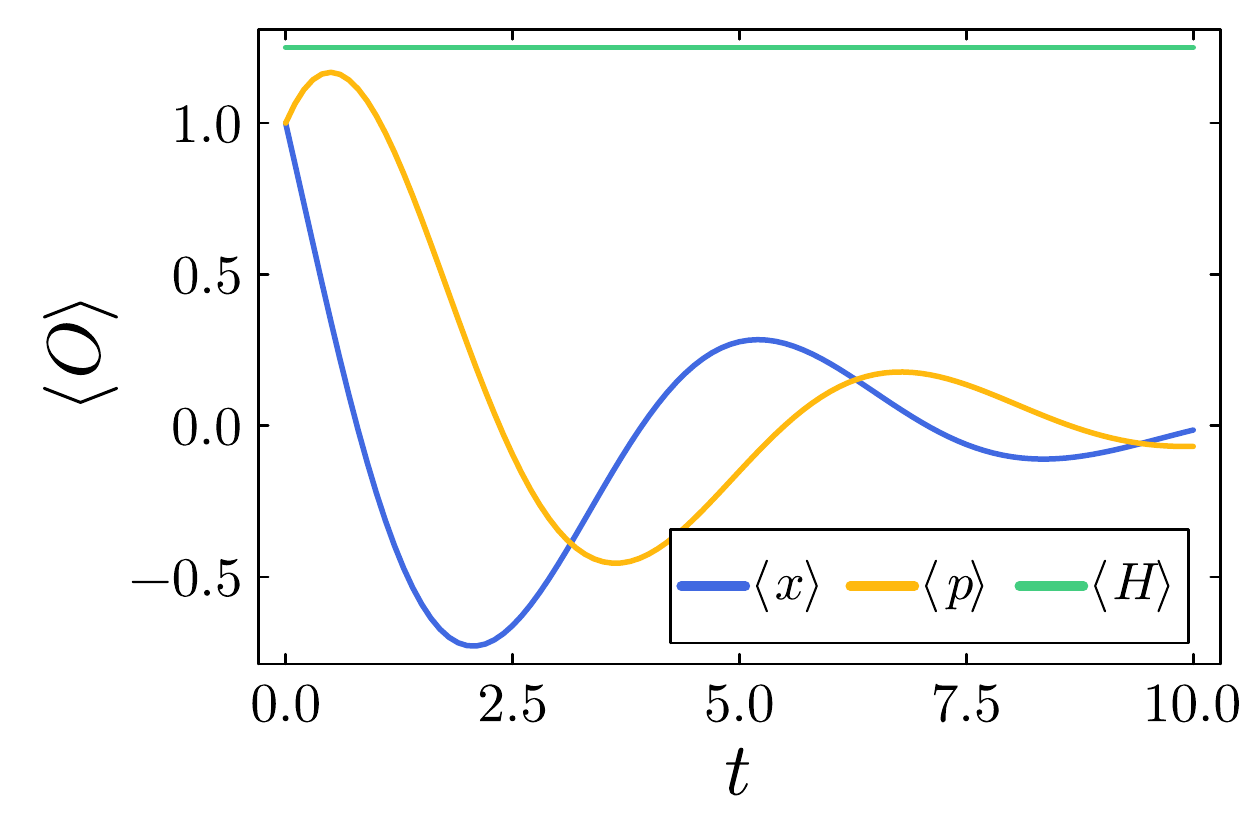}
    \includegraphics[width=0.32\linewidth]{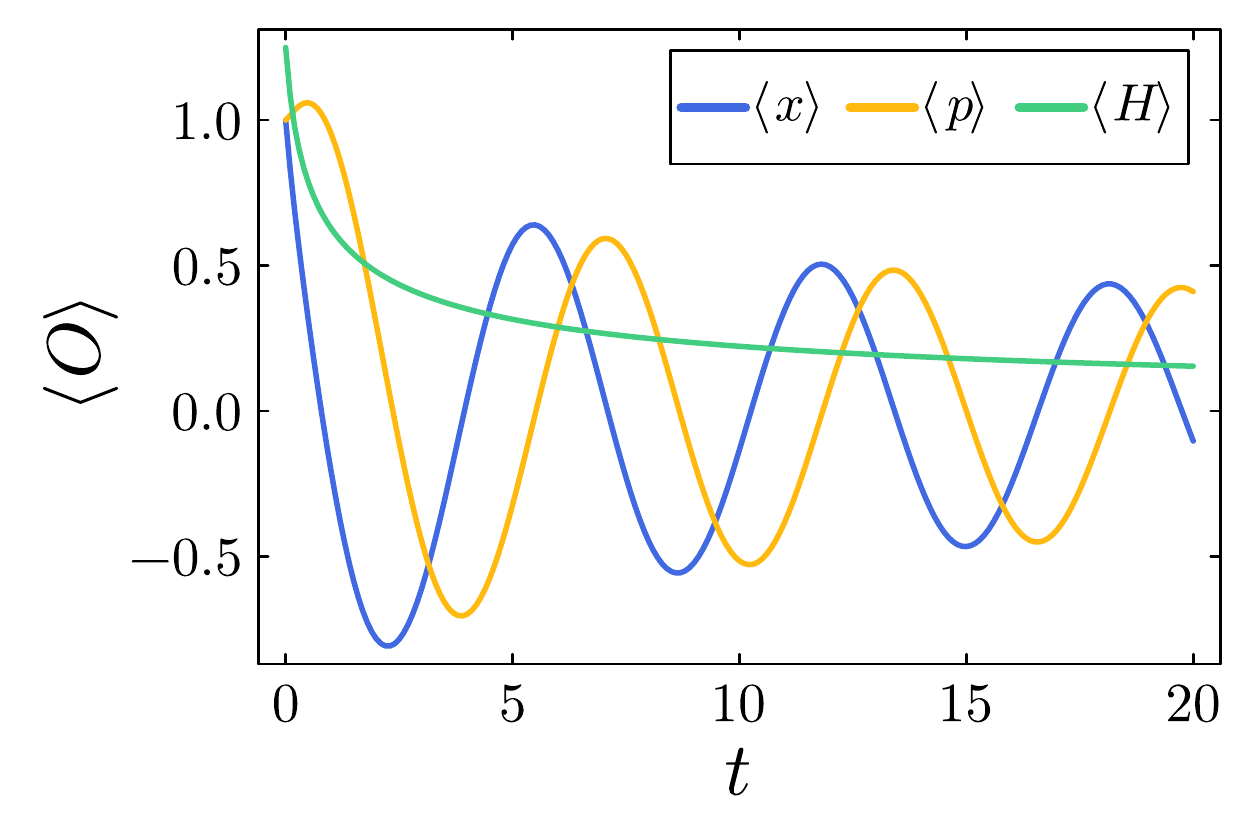}
    \includegraphics[width=0.32\linewidth]{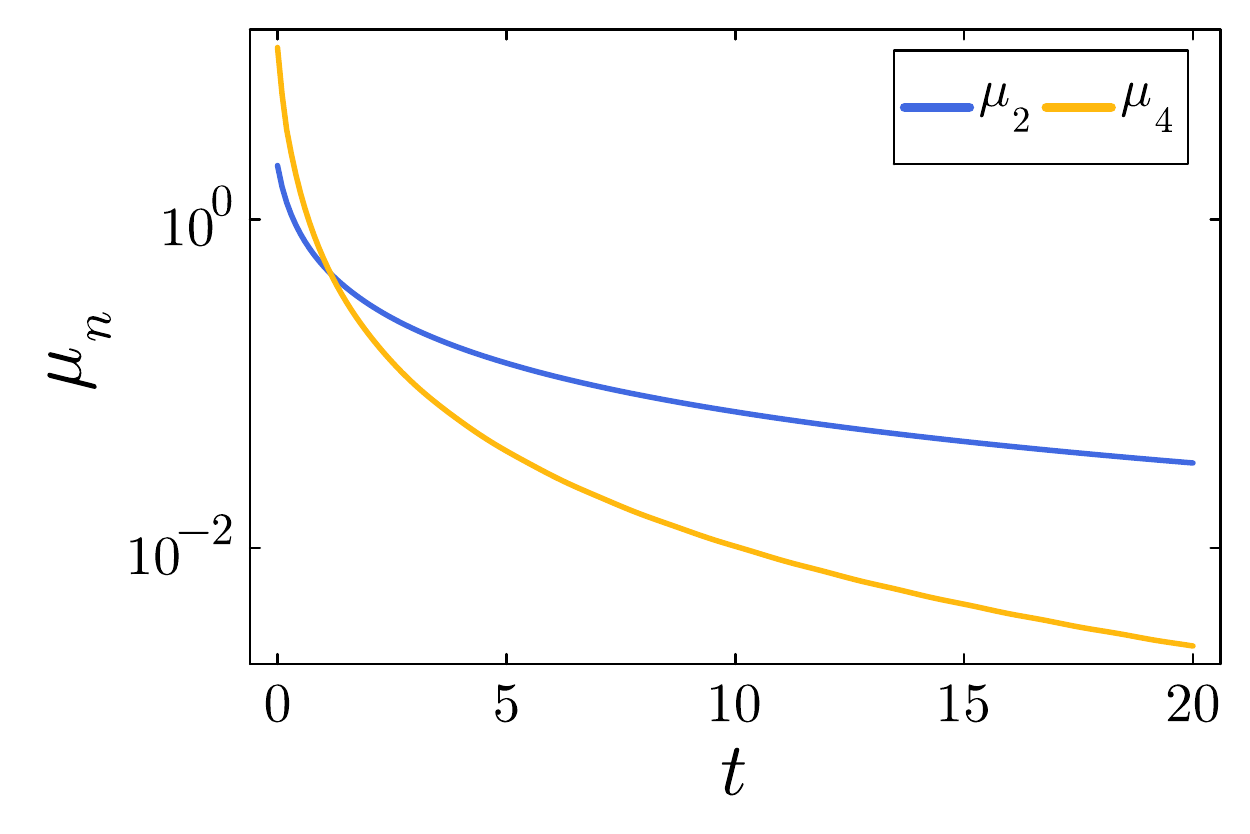}
     \caption{Average position, momentum, and energy of the Wigner function evolved using the semiclassical limit of the master equation with double commutator (left) and double anticommutator (middle). We can also observe the monotonic decrease of energy moments $\mu_2$ and $\mu_4$ (right) for the case of double anticommutator.} 
    \label{fig:avg}
\end{figure*}

We start our analysis by revisiting the harmonic oscillator. It is described by the Hamiltonian
\begin{align}
    H(x,p,t) = \frac{p^2}{2m} + \frac{1}{2}m\omega^2x^2\,. \label{SHO Hamiltonian}
\end{align}
Since the Hamiltonian is quadratic, the Moyal bracket is exactly equal to the Poisson bracket. The time evolution of the Wigner function in the classical limit of energy dephasing is given by replacing $H(x,p,t)$ in Eq. \eqref{double comm evolution equation}, 
\begin{align}
    \partial_t W =& m \omega^2x \,\partial_p W - \frac{p}{m}\, \partial_x W \nonumber \\
    &+ \gamma [m^2 \omega^4x^2 \partial^2_pW + \frac{p^2}{m^2} \partial^2_xW - 2\omega^2 xp\,\partial_x\partial_p W \nonumber \\
    &- \omega^2 p\, \partial_p W -  \omega^2x\,\partial_x W]\,.  \label{HO}
\end{align}
 The numerical solution to the above equation is plotted in Fig.~\ref{fig:1} for a Gaussian initial state. We have also plotted the average position, momentum, and energy of the Wigner function in Fig.~\ref{fig:avg}.

The problem is more transparent in the canonical action-angle variable, where the Hamiltonian can be written as $H  =\omega r^2/2$ with $r^2 = X^2 + P^2$, $X = \sqrt{m \omega}\, x$, and $P = p/\sqrt{m\omega}$. We can define the Wigner function $W(I, \theta)$ in terms of variables  $I = r^2/2$ and $\theta = \tan^{-1}(P/X)$.  The corresponding equation for the Wigner function is
\begin{align}
    \partial_t W (I, \theta) = -\omega\, \partial_\theta W(I, \theta) + \gamma \omega^2\, \partial^2_\theta W (I, \theta) \,.
\end{align}
The two terms describe Hamiltonian advection and angular diffusion (for $\gamma >0$). If we write the Wigner function in terms of its Fourier modes $W(I, \theta, t) = \sum_{k \in \mathbb{Z}} \tilde{W_k}(I,t) e^{ik\theta}$, we can simplify the right-hand side
\begin{align}
    \partial_t \tilde{W_k}(I,t) = (-ik\omega - \gamma\omega^2k^2) \tilde{W_k}(I,t)\,,
\end{align}
which yields
\begin{align}
    \tilde{W_k}(I,t) = \tilde{W_k}(I,0)\exp[(-ik\omega - \gamma \omega^2k^2)t]\,.
\end{align}
Thus, the Wigner function evolves as
\begin{align}
    W(I, \theta,t) = \sum_{k \in \mathbb{Z}} \tilde{W_k}(I,0)e^{ik\theta}\exp[(-ik\omega - \gamma \omega^2k^2)t]\,. 
\end{align}
We observe that each mode rotates separately with an exponential damping proportional to $k^2$. The Wigner function in Fourier space $\tilde{W_k}(I, t) \to 0$ as $t \to \infty$ (for $k \neq 0$). Therefore, the long-time stationary state of the Wigner function is uniform in the angle $\theta$ (only $k=0$ survives) for a fixed value of energy. This is evident in the numerical results shown in Fig. \ref{fig:1}, where the Wigner function forms a uniform ring around the center. The corresponding phenomenon in the quantum regime is the decay of off-diagonal terms $\rho_{mn}=\langle m|\hat{\rho}|n\rangle$ in the density matrix due to energy dephasing. Equation \eqref{ME4ED} can be written in the energy basis of the Hamiltonian as
\begin{align}
    \partial_t \rho_{mn} = -\frac{i}{\hbar} (E_m - E_n)\rho_{mn} - \frac{\gamma}{2} (E_m - E_n)^2 \rho_{mn}\,,
\end{align}
which can be solved to get
\begin{align}
    \rho_{mn}(t) = \rho_{mn}(0) \exp\left(-\frac{i}{\hbar} \Delta_{mn}t - \frac{\gamma}{2} \Delta_{mn}^2t \right)\,,
\end{align}
where $\Delta_{mn} = E_m - E_n$. For the harmonic oscillator, $\Delta_{mn} = \hbar\omega(m-n)$ and the factor of coherence decay matches the classical Fourier-mode decay with $k = m-n$. Parts of this analysis hold true in general for the Hamiltonians depending purely on the action $H(I)$.

In the semi-classical limit, we can also look at the energy marginal 
\begin{align}
    \overline{W}(I,t) = \frac{1}{2\pi} \int_0^{2\pi} W(I, \theta, t)\mathrm{d}\theta. \label{energy marginal}
\end{align}
This quantity is similar to the population in the energy basis given by the diagonal elements of the density matrix, since both describe the probability distribution across different energy values. Differentiating with respect to time, we get
\begin{align}
    \frac{d \overline{W}(I,t)}{dt} &=  \int_0^{2\pi} [-\omega\, \partial_\theta W(I, \theta, t) + \gamma \omega^2\, \partial^2_\theta W (I, \theta, t)]\frac{\mathrm{d}\theta}{2\pi}  \nonumber \\
    &= \frac{1}{2\pi}  \left[ -\omega\, W(I, \theta, t)\bigg|^{2\pi}_0 + \gamma \omega^2\,\partial_\theta W(I, \theta, t)\bigg|^{2\pi}_0\right] \nonumber\\
    &= 0\;,
\end{align}
using the periodicity of the Wigner function $W(I, \theta, t)$ along the angle variable $\theta$. In the quantum limit, we observe a similar behavior as the population in different energy eigenstates remains unchanged in the presence of dephasing. The mean energy $\int I\overline{W}(I,t)\,{\rm d}I$ is also a constant, as can be observed from the numerical simulation in Fig.~\ref{fig:avg}. This is consistent with the invariance of the total energy, as expected from the original master equation.

In Section \ref{SecDAME}, we derived the expression for the classical limit of the double anticommutator term that appears in the dynamics of stochastic non-Hermitian Hamiltonians. We can numerically study the evolution of Wigner functions governed by such equation for the same physical systems. We begin by truncating Eq. \eqref{W evol double AC} to $\mathcal{O}(\hbar^0)$ to study the effect of the balanced gain/loss term in the usual Liouvillian flow induced by Poisson bracket. We have plotted the numerical results for the harmonic oscillator in Fig. \ref{fig:1}.

Before delving into the classical limit of the double anticommutator master equation, we first analyze the quantum phenomenon that Eq. \eqref{DC Master eqn} captures. In the eigenbasis of the Hamiltonian $\hat{H}$, the evolution of the elements of the density matrix is given by
\begin{align}
    \partial_t\rho_{mn} = \frac{E_m - E_n}{i\hbar}\rho_{mn} - \Gamma[(E_m + E_n)^2 - 4 \langle \hat{H}^2 \rangle]\rho_{mn}\,.
\end{align}
Focusing on the diagonal elements, we get the expression
\begin{align}
    \partial_t \rho_{n} = -4\Gamma(E_n^2 - \langle \hat{H}^2 \rangle)\rho_n\,,
\end{align}
where $\hat{\rho}_{nn}(t) \equiv \hat{\rho}_n(t)$. This equation is of the exact same form as Eq. \eqref{Wig Solution}, and its solution is thus
\begin{align}
    \rho_n(t) = \frac{\rho_n(0)e^{-4 \Gamma E_n^2t}}{\sum_{k} \rho_k(0)e^{-4\Gamma E_k^2t}}\,,
\end{align}
We note that, in the large $t$ limit, populations in the higher-energy levels are exponentially suppressed. Thus, Eq. \eqref{DC Master eqn} describes cooling in the quantum regime. The system is driven towards its ground state, provided there is a nonzero overlap between the ground state and the initial state. Otherwise, the system will drive towards the state with the minimum possible energy among those with non-zero overlap. If we instead focus on the off-diagonal elements ($m \neq n$), we get 
\begin{align}
    \rho_{mn}(t) = \frac{\rho_{mn}(0) e^{-\frac{i}{\hbar}(E_m - E_n)t}e^{-\Gamma(E_m + E_n)^2t}}{\sum_{k} \rho_k(0)e^{-4\Gamma E_k^2t}}\,.
\end{align}
Along with the usual oscillation terms with frequency determined by the energy gap $\Delta_{mn} = E_m - E_n$, the damping term induces a decay rate proportional to $(E_m + E_n)^2$. This implies decoherence in the energy eigenbasis, where the elements farther from the diagonal (or, equivalently, the coherent terms involving higher energies) damp faster. 

As before, we can analyze the dynamics of the harmonic oscillator in terms of action-angle variables. The energy marginal, as defined in Eq. \eqref{energy marginal}, satisfies the differential equation
\begin{align}
    \partial_t \overline{W}(I,t) = -4\Gamma (I^2 - \mu_2(t))\overline{W}(I,t)\,,
\end{align}
which can be formally solved to get
\begin{align}
    \overline{W}(I,t) = \frac{\overline{W}(I,0)e^{-4\Gamma I^2 \,t}}{\int \overline{W}(I,0)e^{-4\Gamma I^2 t} \mathrm{d}I}\,.
\end{align}
Notice that the energy marginal is not constant as in the classical limit of the double commutator. Higher energies are exponentially suppressed, and the Wigner function in energy space will concentrate towards the lowest possible energy present in the initial distribution. As the probability distribution shifts towards the lowest energy value, we expect the mean energy $\int I\, \overline{W}(I,t)\,{\rm d}I$ to decay as well. This can be observed in Fig. \ref{fig:avg}.

For the Gaussian distribution chosen as the initial state, we can explicitly calculate the energy marginal. Defining $I_0 = (x_0^2 + p_0^2)/2$ and $\phi = \tan^{-1}(p_0/x_0)$, the initial distribution in terms of action-angle variables reads
\begin{align}
    W(I,0) = \frac{1}{2\pi \sigma^2}\, \exp\left(-\frac{(I + I_0) - \sqrt{II_0} \cos(\theta - \phi)}{\sigma^2}\right)\,.
\end{align}
Plugging this into the expression for the energy marginal, 
\begin{align}
    \overline{W}(I,0) &= \frac{1}{2\pi} \int_0^{2\pi} W(I,0)\mathrm{d}\theta \nonumber\\
    &= \frac{1}{2 \pi \sigma^2}\, \exp\left(-\frac{I + I_0}{\sigma^2}\right)\; \mathcal{I}_0 \bigg(\frac{2 \sqrt{I I_0}}{\sigma^2}\bigg)\,,
\end{align}
where we have used the definition of the modified Bessel function $\int_0^{2\pi}e^{z \cos{u}}{\rm d}u = 2\pi \,\mathcal{I}_0(z)$. The energy of the Gaussian state can be arbitrarily close to zero, which implies that the Wigner function eventually evolves towards the low-energy state, i.e., it forms a peak closer to the center in $(x,p)$ space. We observe this numerically in Fig. \ref{fig:1}. This phenomenon is equivalent to cooling and is consistent with the diagonal elements of the density matrix evolving in the energy basis when the spectrum has nondegenerate $E_n^2$ values.

\subsection{Decoherence and Quantum-Classical Correspondence in Chaotic Models} \label{Deco and QC correspondence}

The system we have studied so far, i.e., the simple harmonic oscillator, provides valuable insights into the effect of the double-bracket structure in the classical limit. However, since the Hamiltonian is quadratic, the Moyal bracket is exactly equal to the Poisson bracket, and there is no quantum effect in the classical phase space in the absence of decoherent terms. Here, we move onto a richer model, namely the driven anharmonic oscillator given by the Hamiltonian
\begin{align}
    \hat{H} = \frac{\hat{p}^2}{2m} - A\hat{x}^2 + B\hat{x}^4+ \kappa \hat{x} \cos{(\omega_d t)}\,.
\end{align}
Due to the presence of a quartic term, there is a $\order{\hbar^2}$ correction to the Moyal bracket given by
\begin{align}
    \{\{H,W\}\} = \{H,W\}_{\rm P} - \hbar^2 B x\, \frac{\partial^3W}{\partial p^3}\,.
\end{align}
Moreover, explicit time dependence in the Hamiltonian makes the system chaotic. The effect of decoherence in this model is studied in \cite{Habib:1998ai}, in the weak-coupling, high-temperature limit of quantum Brownian motion. The authors show that quantum effects are suppressed by decoherence, as the Wigner function remains positive and the expectation values of observables agree when evolved by both the quantum master equation and the classical Fokker-Planck equation. In the semi-classical limit, a driven anharmonic oscillator has also been studied in \cite{PhysRevLett.65.2927} using the Husimi distribution function, where the authors find that the driving term enhances chaos and increases the rate of quantum tunneling between stability tubes present in the classically inaccessible regions of phase space.

Another canonical chaotic model studied in this context is the kicked rotor. Authors in \cite{jensen1990wigner} showed that quantum corrections to the classical motion scale much faster in the chaotic limit of the kicked rotor compared to the case of smaller kick amplitude with quasiperiodic trajectories in phase space. However, in the presence of decoherence, Refs. \cite{Zurek:1994wd, kolovsky1996quantum} argue that higher-order quantum corrections to the Poisson bracket can be ignored for longer times in chaotic systems. It is because diffusive effects prevent the formation of small-scale structures in phase space, thereby setting the time scale for quantum effects to significantly alter the dynamics. 

In the previous section, we noted that the presence of a double anticommutator term in the quantum master equation adds a classical term to the evolution equation for the Wigner function; see Eq. \eqref{cl_eq_DAC}. Here, we describe its effect in the presence of decoherence and the quantum correction term to the Poisson bracket. Consider the 
equation
\begin{align}
    \partial_t W = \{&H,W\}_{\rm P} + \gamma \{H, \{H,W\}_{\rm P} \}_{\rm P} \nonumber \\&-4\Gamma (H^2 - \langle H^2 \rangle)W- \hbar^2 B x\, \frac{\partial^3W}{\partial p^3}\,. \label{Quantum_PDE}
\end{align}

We have thus far neglected higher-order contributions arising from the double-bracket structures. To second leading order, these terms are given explicitly in Eq.~\eqref{evol W} for the double-commutator case and in Eq.~\eqref{W evol double AC} for the double-anticommutator case, and can in principle be evaluated analytically for the driven anharmonic oscillator Hamiltonian. However, since our analysis focuses on the semiclassical regime of weak dissipative coupling ($\hbar \lesssim 0.5$, $\gamma \sim \Gamma \lesssim 0.1$), corrections of order $\mathcal{O}(\hbar^2\gamma)$ and $\mathcal{O}(\hbar^2\Gamma)$ remain negligibly small throughout and can therefore be safely omitted.

\begin{figure*}[ht]
    \centering
    \includegraphics[width=0.4\linewidth]{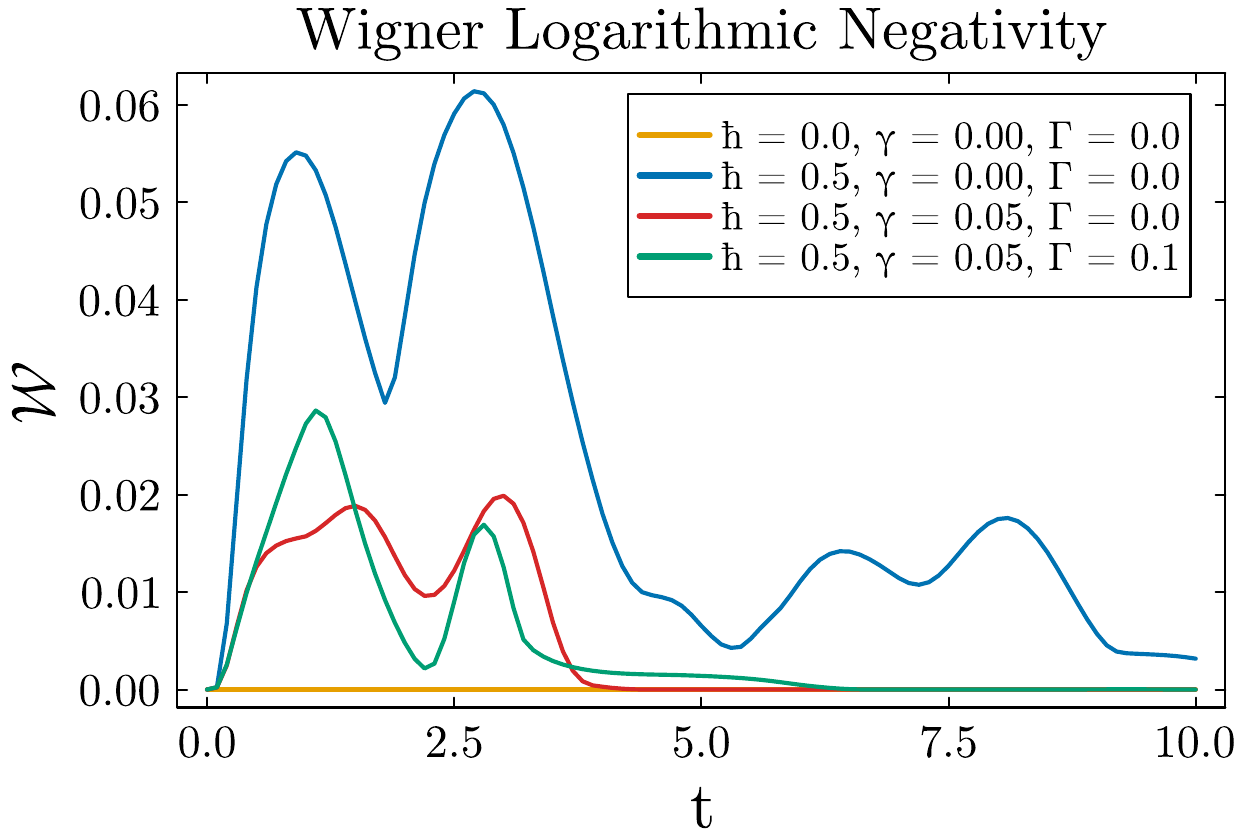}
    \includegraphics[width=0.4\linewidth]
    {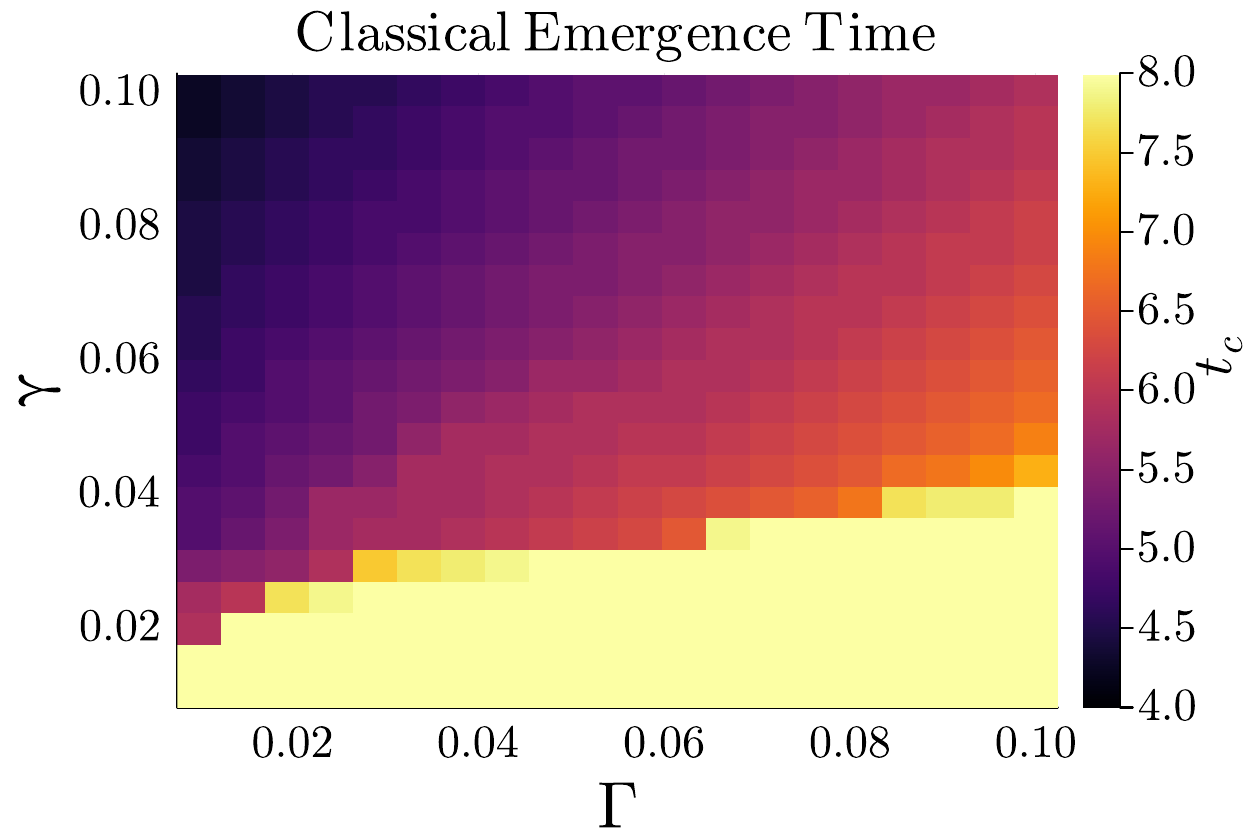}
    \caption{Wigner logarithmic negativity $\mathcal{W}$ and time of classical emergence $t_c$ for different parameter values. Physical parameters in the Hamiltonian are set to $m = 1.0$, $A = 1.0$, $B = 0.1$, $\kappa = 0.2$, and $\omega_d = 1.0$. The initial state is taken to be a Gaussian wavepacket with minimum uncertainty, centered near the right minimum at $x_0 = 2.19$.}
    \label{fig:Wig_neg}
\end{figure*}

For an initial Gaussian wavepacket in phase space, the $\order{\hbar^2}$ quantum term is expected to introduce negative values in the Wigner function, which is an indicator of non-classical behavior \cite{lutkenhaus1995nonclassical, Kenfack:2004ges}. It is known that the decoherence terms suppress these negative values, indicating the emergence of classicality \cite{strandberg2019numerical}. We next use a resource monotone called Wigner Logarithmic Negativity $\mathcal W$, defined as
\begin{align}
    \mathcal{W} = \log\bigg(\int |W(x,p)|\mathrm{d}x\mathrm{d}p\bigg)\;,
\end{align}
to capture the negativity in the Wigner function. This probe captures the ``quantum-ness'' of the system and is closely related to its stabilizerness~\cite{goto2022probing}.  For a normalized Wigner function $W(x,p)$, $\mathcal{W}$ is positive iff there exists a finite region $\Omega$ such that $W(x,p)\vert_{x,p \in \Omega} < 0$. Figure \ref{fig:Wig_neg} shows how the Wigner logarithmic negativity evolves for different parameter values. As expected, there are no negative values without the quantum correction, and the decoherence term eventually suppresses the negative values. We define the time of classical emergence $t_c$ as the instant when the total negative area of the Wigner function reaches zero; see also \cite{Xu25}. Figure \ref{fig:Wig_neg} also shows $t_c$ for different values of the parameters $\gamma$ and $\Gamma$.
\begin{figure}[ht]
    \centering
    \includegraphics[width=1.0\linewidth]{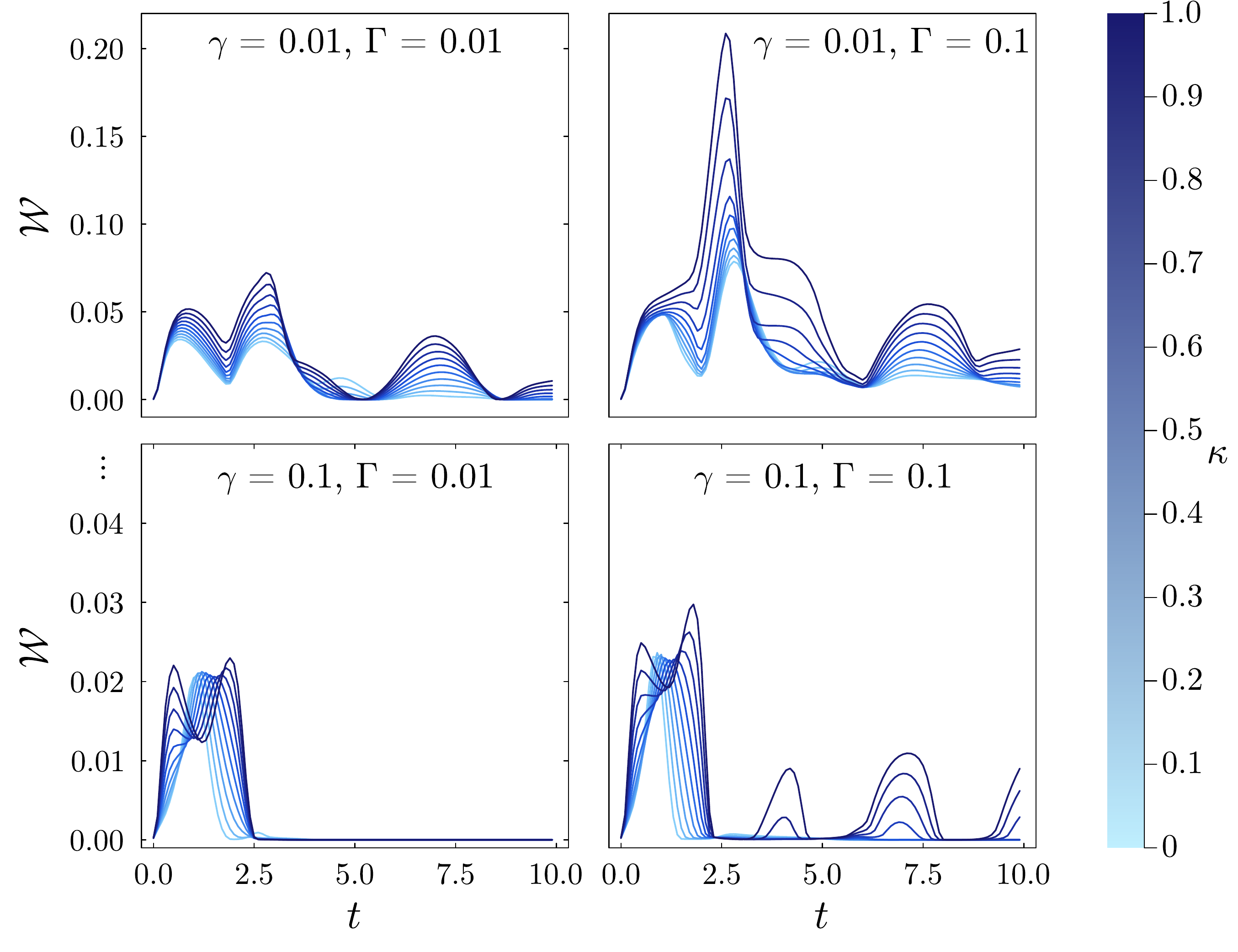}
    \caption{Wigner logarithmic negativity $\mathcal{W}$ for different strengths of the nonlinearity parameter $\kappa \in [0,1]$ plotted for different decoherence strengths $\gamma$ and $\Gamma$ using the Gaussian initial state in the driven anharmonic oscillator.}
    \label{fig:Wig_neg_Lambda}
\end{figure}
We observe that the presence of the classical term from the anticommutator part delays the emergence of classicality in the Wigner function. From the differential equation, it follows that the classical term amplifies or dampens the Wigner function depending on the relative sign of $(H^2 - \langle H^2 \rangle$) at each point in the phase space. It acts as a multiplicative factor and thus cannot generate negative values in the Wigner function. However, it can amplify the negative values introduced by the quantum contribution and help to maintain them longer against decoherence. The balanced gain/loss contribution in the evolution equation provides a mechanism to delay the quantum-to-classical transition in phase space. 
The negativity of the Wigner function can also be used as a resource for quantum computation \cite{pashayan2015estimating, albarelli2018resource}. Prolonged negativity indicates that the anti-Hermitian noise in the Hamiltonian (which results in the double anticommutator term in the quantum master equation) can act as a resource for quantum advantage. An experimental approach to measuring the Wigner function, as demonstrated in \cite{smithey1993measurement, nogues2000measurement}, can be used to verify the numerical observation. In Fig. \ref{fig:Wig_neg_Lambda}, we plot the Wigner logarithmic negativity $\mathcal{W}$ for different strengths of the nonlinear term parameterized by $\kappa$. We note that the peak of $\mathcal{W}$ is higher on average for a larger coupling strength. Increasing the strength of the decoherence term with $\gamma$ (moving below the grid) suppresses the Wigner negativity. However, increasing $\Gamma$ (moving towards the right in the grid) enhances the negative regions and helps them to last longer against decoherence.       

The above analysis can be extended to a quantum (non-coherent) initial state. Consider Schr\"{o}dinger cat states, which are linear superpositions of classically distinguishable states. They can be realized in multiple physical systems. For instance, in quantum optics, using coherent states
\begin{align}
\ket{\alpha} = \exp\left(-\frac{1}{2}|\alpha|^2\right) \sum_{n=0}^\infty \frac{\alpha^n}{\sqrt{n!}}\ket{n}\,,
\end{align}
where $\ket{n}$ is the $n$-th eigenstate of the quantum harmonic oscillator,  a cat state can be defined as
\begin{align}
    \ket{\psi_{\text{cat}}} = \mathcal{N}\left(\ket{\alpha} + e^{i\phi}\ket{-\alpha}\right)\,.
\end{align}
Here, $\phi$ captures the relative phase between the coherent states, and some special cases include $\phi = 0$ (even coherent states), $\phi = \pi$ (odd coherent states), and $\phi = \pi/2$ (Yurke-Stoler states)  \cite{gerry1997quantum}. These states have been experimentally realized on various platforms \cite{Bild23, vlastakis2013deterministically, he2023fast, wang2016schrodinger, leghtas2015confining, yu2025schrodinger}, and are a useful resource for understanding quantum-classical boundaries and for quantum information processing applications. In terms of the Wigner function, the cat state is defined as 
\begin{align}
    W_{\text{cat}} = \mathcal{N} &\big\{\exp[-2((x - \alpha)^2 + p^2)] \nonumber \\&+ \exp[-2((x + \alpha)^2 + p^2)]\nonumber \\
     &+ 2 \exp[-2(x^2 + p^2)]\cos{(4\alpha\, p - \phi)}\big\}\,,
\end{align}
with $\mathcal{N} = 1/(\pi[1 + \cos{\phi\exp{-2\alpha^2}}])$ as the normalization constant \cite{gerry1997quantum}. The first two terms describe classical Gaussian states centered at $(\alpha,0)$ and $(-\alpha,0)$, respectively. The final term captures quantum interference and introduces negative regions in phase space. We set the phase parameter $\phi = 0$ to describe an even parity coherent state.
\begin{figure*}[htbp]
    \centering
    \includegraphics[width=0.32\linewidth]{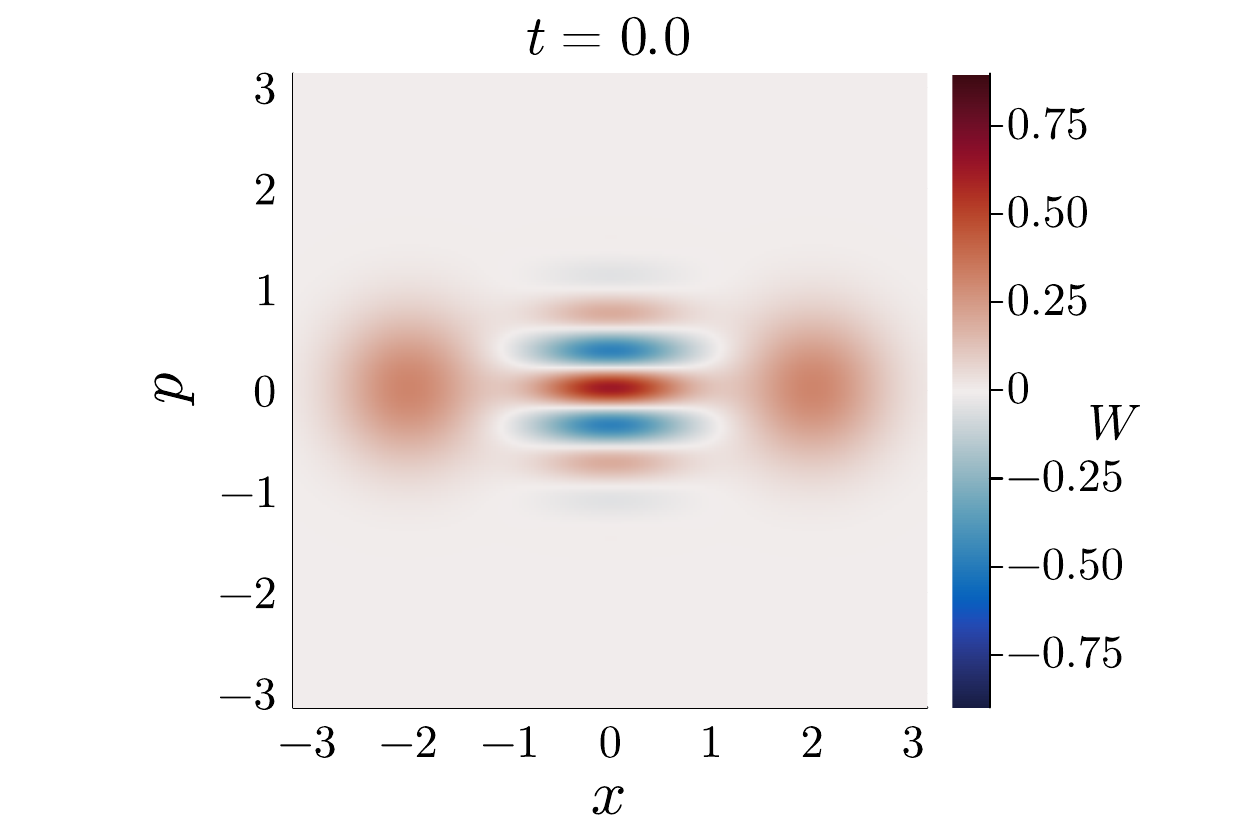}
    \includegraphics[width=0.32\linewidth]{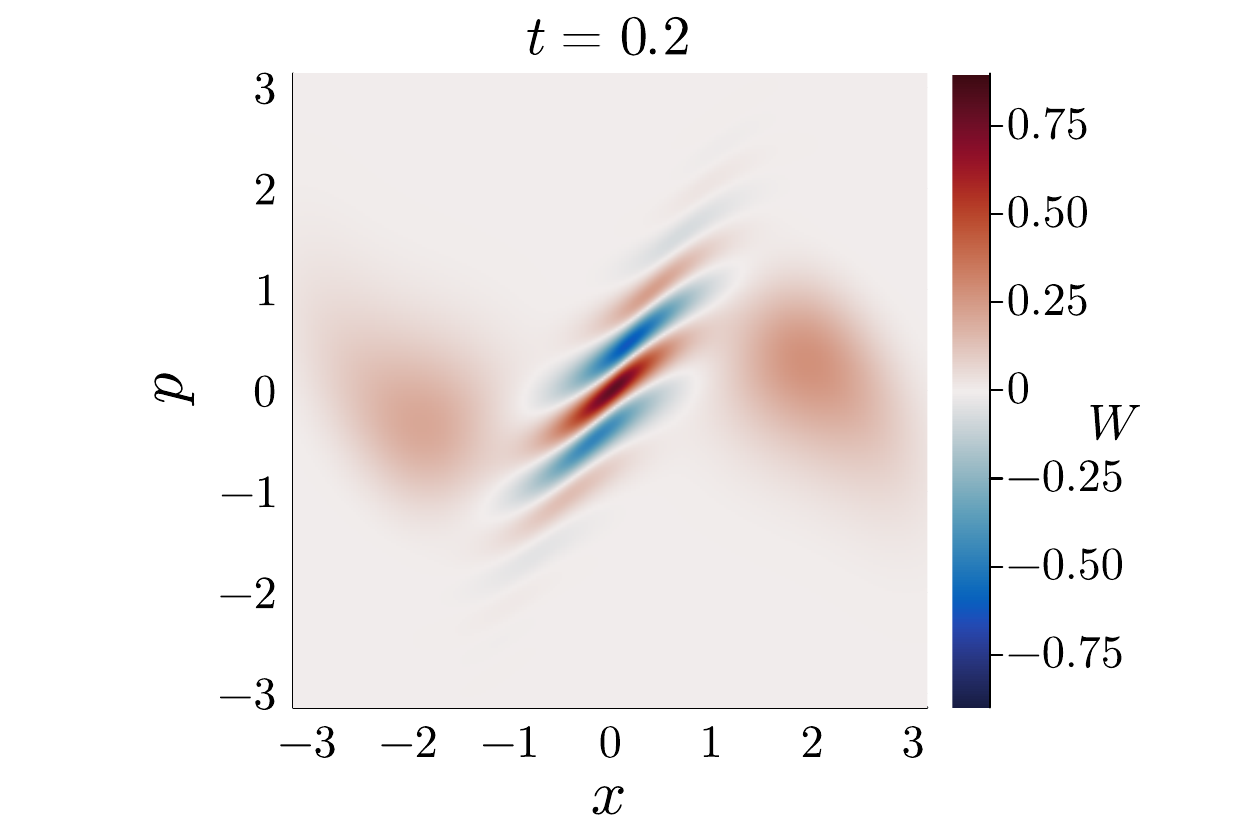}
    \includegraphics[width=0.32\linewidth]{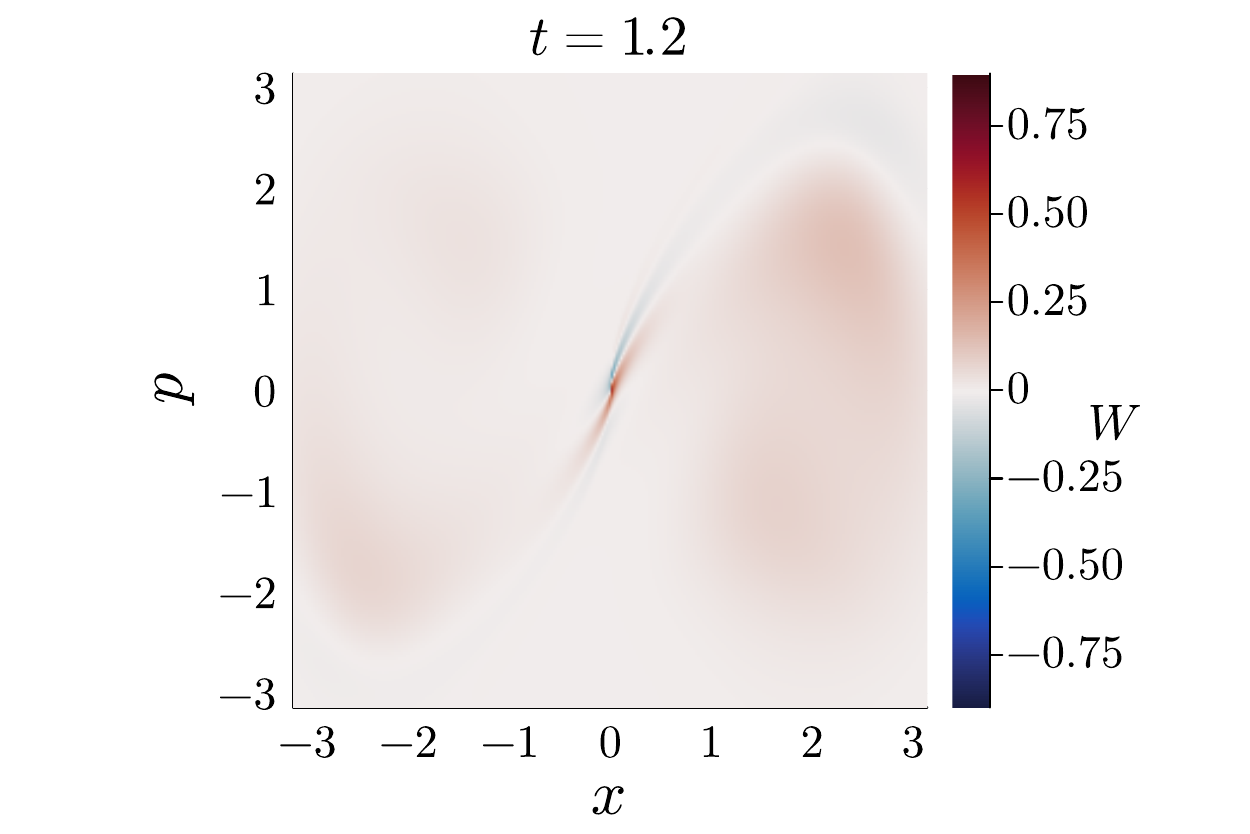}
    \caption{Evolution of the Wigner function of an initial cat state $W_{\text{cat}}$ under the driven anharmonic oscillator potential with parameter values $\gamma = 0.05$, $\Gamma = 0.1$, and $\kappa = 0.2$.}
    \label{fig: Cat state Wigner function}
\end{figure*}

\begin{figure}[h]
    \centering
    \includegraphics[width=\linewidth]{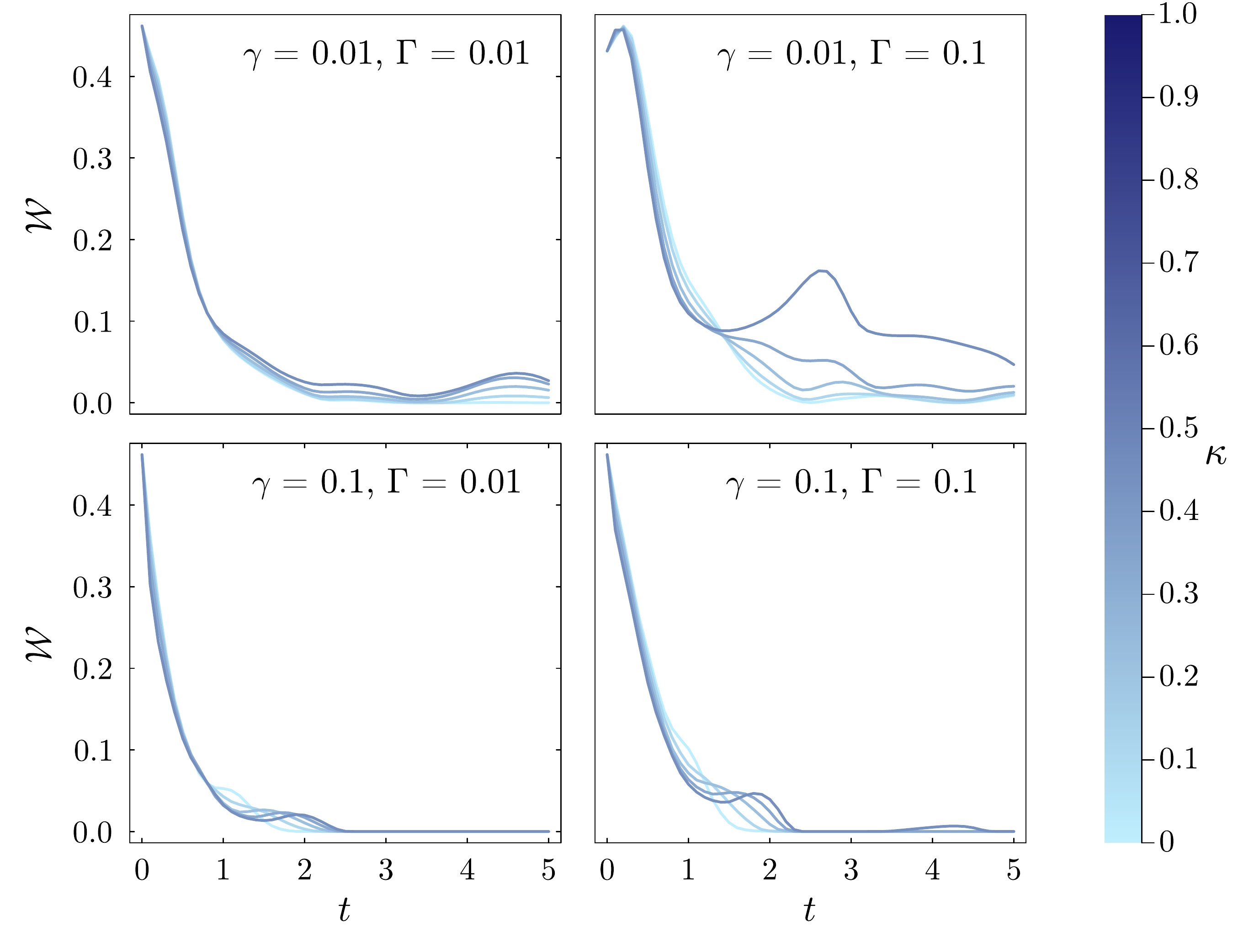}
    \caption{Wigner Logarithmic Negativity $\mathcal{W}$ for different parameter values of the anharmonic oscillator and an initial cat state.}
    \label{Wig Log Neg CAT}
\end{figure}
In Fig. \ref{fig: Cat state Wigner function}, we present the evolution of the Wigner function obtained by solving the dynamical equations \eqref{Quantum_PDE} with the Hamiltonian for a driven Anharmonic oscillator and $ W_{\text{cat}} $ as the initial state. Negative regions are highlighted in blue. Similarly, Fig. \ref{Wig Log Neg CAT} shows the Wigner logarithmic negativity $\mathcal{W}$ as a function of time and captures the effect of decoherence and anti-dephasing term at various strengths. Across different values of the chaos parameter $\kappa$, we observe that increasing $\Gamma$ for fixed decoherence strength $\gamma$ (moving right along the grid) enhances the Wigner negativity and extends it for longer times. Similarly, increasing the decoherence strength (moving below the grid) suppresses the negative regions, indicating the emergence of classical behavior.

\section{Beyond double bracket master equations: Spectral Filtering and higher-order nested brackets}\label{SecGenME}

Before closing, we discuss examples of master equation involving higher-order nested brackets. 
Numerical methods for studying the spectral properties of many-body systems often employ filters that select a subset of the spectrum (eigenvalue filters) or a frequency range (frequency filters). Frequency filters can be associated with Liouvillian deformations of energy dephasing, and can be described in terms of master equations with higher-order nested brackets \cite{MatsoukasRoubeas23,apollo2023unitarity}. Specifically, given the Hamiltonian $\hat{H}=\sum_nE_n|n\rangle\langle n|$, we consider a frequency filter $w(x)\geq 0$, satisfying $w(x)=w(-x)$, such as that used in the study of the filtered spectral form factor ${\rm SFF}_{w}(t)=\frac{1}{d^2}\sum_{nm}w(E_n-E_m)e^{-it(E_n-E_m)}$, where $d$ is the Hilbert space dimension.
The spectral form factor ${\rm SFF}_{w}(t)$ can be understood as the Uhlmann fidelity between the quantum state $\hat{\rho}(0)=\frac{1}{d}\sum_{nm}|n\rangle\langle m|$ and its time evolution \cite{Xu21SFF,Cornelius21}
\begin{equation}
\hat{\rho}(t)=\frac{1}{d}\sum_{nm}|n\rangle\langle m|e^{-it(E_n-E_m)}e^{\chi(t) G(E_n-E_m)}\,,
\end{equation}
in terms of the auxiliary function $G(x)=\log w(x)$. The real function $\chi(t)$ satisfies $\chi(0)=0$. Frequent choices for it involve the Heaviside step function $\chi(t)=\gamma\Theta(t)$ with some $\gamma<0$ and the linear function $\chi(t)=\gamma t$, which arises naturally in certain non-Hermitian quantum evolutions \cite{apollo2023unitarity}. In particular, the case of energy dephasing is associated with $\chi(t)=\gamma t$ and $W(x)=-x^2$, i.e., for a Gaussian frequency filter $w(x)$.  
The master equation describing the action of a general frequency filter can thus be written as \cite{apollo2023unitarity}
\begin{equation}
\frac{d}{dt}\hat{\rho}=-i[\hat{H},\hat{\rho}]+\dot{\chi}(t)\sum_{n=0}^\infty \frac{G^{(2n)}(0)}{(2n)!}[\hat{H},\hat{\rho}]_{2n}\,,
\end{equation}
in terms of the $n$-th order nested commutator $[\hat{H},\hat{\rho}]_{n}=[\hat{H},[\hat{H},\hat{\rho}]_{n-1}]$. 
The form of the dissipator reflects the Taylor series expansion of the symmetric function 
$G(z)=\sum_{n=0}^\infty\frac{G^{(2n)}(0)}{(2n)!}z^{2n}$.
This is an instance of a master equation involving higher-order nested commutators. 

Motivated by this, we consider extending such a master equation to a system of continuous variables. In phase space, each term in the dissipator is found by the map 
\begin{equation}
[\hat{H},\hat{\rho}]_{2n}\mapsto \left(H\frac{2}{\hbar}\sin\left(\frac{\hbar}{2}\Lambda\right)\right)^{2n}W\,. 
\end{equation}
Alternatively, using the Taylor series of $G(z)$, the equation of motion of the Wigner function can be compactly written as
\begin{equation}
\frac{d}{dt}W=\frac{2}{\hbar}H \sin \left(\frac{\hbar}{2}\Lambda\right)W+\dot{\chi}(t)G\left[H\frac{2}{\hbar}\sin \left(\frac{\hbar}{2}\Lambda\right)\right] W\,.
\end{equation}
To leading order in the $\hbar$-expansion, one obtains the classical equation of motion
\begin{eqnarray}
\frac{d}{dt}W &=&\{H,W\}_{\rm P}+\dot{\chi}(t)G\left(H\Lambda\right) W\\
&=&\{H,W\}_{\rm P}+\dot{\chi}(t)\sum_{n=0}^\infty \frac{G^{(2n)}(0)}{(2n)!}\{H,W\}_{\rm P}^{2n}\,,\nonumber
\end{eqnarray}
in terms of the $n$-th order nested Poisson bracket $\{H,W\}_{\rm P}^n$, with $\{H,W\}_{\rm P}^1=\{H,W\}_{\rm P}$, $\{H,W\}_{\rm P}^2=\{H,\{H,W\}_{\rm P}\}_{\rm P}$, and so on. The evolution remains deterministic, and the locality of the dissipation is reduced with increasing order of the nested commutators.

Similarly, one can justify master equations with higher-order nested anticommutators. Consider the spectral filtering of the density matrix
\begin{equation}
\hat{\rho}(t)=\frac{1}{N(t)}\sum_{nm}\hat{\rho}_{nm}(0)|n\rangle\langle m|e^{-it(E_n-E_m)}e^{\chi(t) G(E_n+E_m)}\,,\label{rhotna}
\end{equation}
with the normalization $N(t)=\tr[\hat{\rho}(0)e^{\chi(t)G(2\hat{H})}]$.
Eigenvalue filters \cite{apollo2023unitarity} are associated with the choice $G(E_n+E_m)=G(E_n)+G(E_m)$.
It can be shown that the evolution (\ref{rhotna}) is associated with the master equation
\begin{eqnarray}
\frac{d}{dt}\hat{\rho}&=&-i[\hat{H},\hat{\rho}]\\
 &+& \dot{\chi}(t)\sum_{n=0}^\infty \frac{G^{(2n)}(0)}{(2n)!}\left[\{\hat{H},\hat{\rho}\}_{2n}-2^{2n}\Tr(\hat{H}^{2n}\hat{\rho})\hat{\rho}\right]\,,  \nonumber  
\end{eqnarray}
where the $n$-th order nested anticommutator obeys $\{\hat{H},\hat{\rho}\}_n=\{\hat{H},\{\hat{H},\hat{\rho}\}_{n-1}\}$.

To derive the classical limit, we explore the phase-space representation of the master equation, which can be obtained by noting that the nested anticommutators map as (see Eq. \eqref{symm Moyal Bracket})
\begin{equation}
\{\hat{H},\hat{\rho}\}_{2n}\mapsto \left(2H\cos\left(\frac{\hbar}{2}\Lambda\right)\right)^{2n}W=:\{\{H,W\}\}_{+}^{2n}\,,
\end{equation}
where we have defined the nested symmetric Moyal bracket $\{\{H,W\}\}_{+}^{n}=\{\{H,\{\{H,W\}\}_{+}^{n-1}\}\}$.
To leading order in the $\hbar$-expansion, $\{\{H,W\}\}_{+}^{2n}=(2H)^{2n}W$.
Likewise, 
\begin{eqnarray}
\tr(\hat{H}^{2n}\hat{\rho})\mapsto\langle H\star H\star\cdots\star H\rangle=:\langle H^{\star 2n}\rangle\,. 
\end{eqnarray}
The Wigner function evolves as
\begin{eqnarray}
\frac{d}{dt}W&=&\{\{H,W\}\}\\
&+& \dot{\chi}(t)\sum_{n=0}^\infty \frac{G^{(2n)}(0)}{(2n)!}\left[\{\{H,W\}\}_{+}^{2n}-2^{2n}\langle H^{\star 2n}\rangle W\right]\,.  \nonumber  
\end{eqnarray}
Given (\ref{clstareq}), the classical equation of motion reads
\begin{eqnarray}
\frac{d}{dt}W&=&\{H,W\}_{\rm P}\\
&+& \dot{\chi}(t)\sum_{n=0}^\infty \frac{G^{(2n)}(0)}{(2n)!}\left[2^{2n}(H^{2n}-\langle H^{2n}\rangle) W\right]\,,  \nonumber  
\end{eqnarray}
which provides an instance of a master equation with higher-order nested Poisson brackets associated with eigenvalue filtering.

\section{Discussion and conclusions}
Double-bracket equations have widespread applications in physics. By resorting to a phase-space formulation, we have analyzed the classical limit of double-bracket equations via an $\hbar$-expansion. 

Specifically, we have considered nonunitary evolutions describing energy dephasing, in which the dissipator is given by a double commutator with the system Hamiltonian.
We have also considered the trace-preserving evolution in which the dissipator involves a double anticommutator, which is the effective nonlinear evolution that emerges as the noise-average dynamics of a stochastic non-Hermitian system.
Both kinds of evolution admit a description in terms of gradient flows. 
We have illustrated the semiclassical evolution in these cases for a simple harmonic oscillator as well as for a driven anharmonic oscillator, which reflects the interplay of decoherence, dissipation, and chaos.

Motivated by the use of filters in the study of the spectral properties in many-body physics, we have also introduced the associated quantum master equations, phase-space dynamical equations and classical equations of motion involving nested brackets (commutators or anticommutators),  Moyal brackets (antisymmetric as well as their symmetric extension), and Poisson brackets, respectively. 

The results presented here indicate that phase-space methods provide a natural and versatile framework for studying nonunitary quantum dynamics generated by double and nested brackets. The emergence of gradient-flow structures and well-defined classical limits shows that these evolutions admit a clear geometric interpretation beyond the strictly quantum regime. Extending this approach to interacting many-body systems, nonlinear filtering techniques, and other instances of quantum chaos is a promising direction for future work, with potential applications to decoherence control and to the semiclassical analysis of spectral filtering techniques. 

\acknowledgements
It is a pleasure to acknowledge discussions with \'Iñigo L. Egusquiza and Zhenyu Xu. 
This work is supported by the  Luxembourg National Research Fund under Grant No. C24/MS/18940482/STAOpen. 

\par \medskip 
\textbf{Data Availability:} Codes for numerical simulations are available in this \href{https://github.com/A-Wenju/DecoQCL#}{Github repository}.

\appendix 

\section{Hamiltonian flow solution for single particle dynamics} \label{Appendix Heat Kernel}
For the single-particle one-dimensional case, 
\begin{align}
    H &= \frac{p^2}{2 m} + V(x)\,,\\
    \mathcal{L} &= \frac{p}{m}\partial_x - V'(x)\partial_p\,.
\end{align}
Hamilton's equations read
\begin{align}
    \frac{d x}{d u} = \frac{\partial H}{\partial p} \;\;,\;\; \frac{d p}{d u} = -\frac{\partial H}{\partial x},
\end{align}
where the time variable along the path is given by $u$. The solutions of these equations describe a trajectory, also known as Hamiltonian or Liouville flow, $\phi_{u}(x,p) \equiv (x(u),p(u))$ with initial condition $\phi_0 (x,p) \equiv (x,p)$. The time evolution of any function of this flow $f(\phi_u (x,p))$ is given by
\begin{align}
    \frac{d f}{d u} &= \frac{\partial f}{\partial x}\frac{d x}{d u} +  \frac{\partial f}{\partial p}\frac{d p}{d u}\notag\\
    &= \frac{\partial f}{\partial x}\frac{\partial H}{\partial p} -  \frac{\partial f}{\partial p}\frac{\partial H}{\partial x}= \{H,f\}_{\rm P}\,.
\end{align}
\pagebreak
Thus, the dynamical equation becomes
\begin{align}
    \frac{d}{d u}f(\phi_u (x,z)) = \mathcal{L}f(\phi_u (x,z))\,.
\end{align}
Hence, the action of $\mathcal{L}$ on the function $f$ can be treated as taking a derivative with respect to $u$. Now, the full solution for $f(t)$, in the presence of the $\mathcal{L}^2$ term is given by
\begin{align}
    f(t) = e^{\gamma t \mathcal{L}^2}f(\phi_{u}(x,z))\Big\vert_{u =t} = e^{\gamma t \partial^2_{u}}f(\phi_{u}(x,z))\Big\vert_{u =t}\,.
\end{align}
Therefore, the operator $\partial^2_u$ is the effect of the $\mathcal{L}^2$ term. This is simply the Laplace-Beltrami operator in $1$-dimension, and therefore the resulting solution is described by the heat kernel as follows
\begin{align}
    f(t) = \frac{1}{\sqrt{2\pi \gamma t}}\int_{-\infty}^{\infty}\exp\left[-\frac{(t - u)^2}{2 \gamma t}\right]f(\phi_u(x,p))\mathrm{d}u\,.
\end{align}
This solution is used to evaluate the Wigner function of the simple harmonic oscillator under the action of the double bracket.

\bibliography{DecoQCl_v7}

@book{BP02,
  title={The Theory of Open Quantum Systems},
  author={Breuer, H. P.  and Petruccione, F.},
  isbn={9780198520634},
  year={2002},
  publisher={Oxford University Press}
}

@Article{Lindblad76,
author="Lindblad, G.",
title="On the generators of quantum dynamical semigroups",
journal="Communications in Mathematical Physics",
year="1976",
month="Jun",
day="01",
volume="48",
number="2",
pages="119--130",
issn="1432-0916",
doi="10.1007/BF01608499",
url="https://doi.org/10.1007/BF01608499"
}

@article{egusquiza1999quantum,
  title = {Quantum evolution according to real clocks},
  author = {Egusquiza, I. L. and Garay, Luis J. and Raya, Jos\'e M.},
  journal = {Phys. Rev. A},
  volume = {59},
  issue = {5},
  pages = {3236--3240},
  numpages = {0},
  year = {1999},
  month = {May},
  publisher = {American Physical Society},
  doi = {10.1103/PhysRevA.59.3236},
  url = {https://link.aps.org/doi/10.1103/PhysRevA.59.3236}
}

@article{egusquiza2003real,
  title = {Real clocks and the {Z}eno effect},
  author = {Egusquiza, I. L. and Garay, Luis J.},
  journal = {Phys. Rev. A},
  volume = {68},
  issue = {2},
  pages = {022104},
  numpages = {10},
  year = {2003},
  month = {Aug},
  publisher = {American Physical Society},
  doi = {10.1103/PhysRevA.68.022104},
  url = {https://link.aps.org/doi/10.1103/PhysRevA.68.022104}
}

@article{chenu2017quantum,
  title = {Quantum Simulation of Generic Many-Body Open System Dynamics Using Classical Noise},
  author = {Chenu, A. and Beau, M. and Cao, J. and del Campo, A.},
  journal = {Phys. Rev. Lett.},
  volume = {118},
  issue = {14},
  pages = {140403},
  numpages = {6},
  year = {2017},
  month = {Apr},
  publisher = {American Physical Society},
  doi = {10.1103/PhysRevLett.118.140403},
  url = {https://link.aps.org/doi/10.1103/PhysRevLett.118.140403}
}

@article{pablo2025quantum,
  title = {Quantum Dynamics with Stochastic Non-{H}ermitian {H}amiltonians},
  author = {Martinez-Azcona, Pablo and Kundu, Aritra and Saxena, Avadh and del Campo, Adolfo and Chenu, Aur\'elia},
  journal = {Phys. Rev. Lett.},
  volume = {135},
  issue = {1},
  pages = {010402},
  numpages = {6},
  year = {2025},
  month = {Jul},
  publisher = {American Physical Society},
  doi = {10.1103/5ksl-tjjm},
  url = {https://link.aps.org/doi/10.1103/5ksl-tjjm}
}

@article{xu2019extreme,
  title = {Extreme Decoherence and Quantum Chaos},
  author = {Xu, Zhenyu and Garc\'{\i}a-Pintos, Luis Pedro and Chenu, Aur\'elia and del Campo, Adolfo},
  journal = {Phys. Rev. Lett.},
  volume = {122},
  issue = {1},
  pages = {014103},
  numpages = {6},
  year = {2019},
  month = {Jan},
  publisher = {American Physical Society},
  doi = {10.1103/PhysRevLett.122.014103},
  url = {https://link.aps.org/doi/10.1103/PhysRevLett.122.014103}
}

@book{nielsen2010quantum,
  title     = {Quantum Computation and Quantum Information},
  author    = {Nielsen, Michael A. and Chuang, Isaac L.},
  year      = {2010},
  publisher = {Cambridge University Press},
  edition   = {10th Anniversary Edition},
  address   = {Cambridge, UK},
  isbn      = {9781107002173}
}

@article{Kiely2017,
  title = {Effect of {P}oisson noise on adiabatic quantum control},
  author = {Kiely, A. and Muga, J. G. and Ruschhaupt, A.},
  journal = {Phys. Rev. A},
  volume = {95},
  issue = {1},
  pages = {012115},
  numpages = {11},
  year = {2017},
  month = {Jan},
  publisher = {American Physical Society},
  doi = {10.1103/PhysRevA.95.012115},
  url = {https://link.aps.org/doi/10.1103/PhysRevA.95.012115}
}

@article{Smith2018,
doi = {10.1088/1367-2630/aa9cd6},
url = {https://doi.org/10.1088/1367-2630/aa9cd6},
year = {2018},
month = {jan},
publisher = {IOP Publishing},
volume = {20},
number = {1},
pages = {013008},
author = {Smith, Andrew and Lu, Yao and An, Shuoming and Zhang, Xiang and Zhang, Jing-Ning and Gong, Zongping and Quan, H T and Jarzynski, Christopher and Kim, Kihwan},
title = {Verification of the quantum nonequilibrium work relation in the presence of decoherence},
journal = {New Journal of Physics},
}

@article{Gluza24,
  doi = {10.22331/q-2024-04-09-1316},
  url = {https://doi.org/10.22331/q-2024-04-09-1316},
  title = {Double-bracket quantum algorithms for diagonalization},
  author = {Gluza, Marek},
  journal = {{Quantum}},
  issn = {2521-327X},
  publisher = {{Verein zur F{\"{o}}rderung des Open Access Publizierens in den Quantenwissenschaften}},
  volume = {8},
  pages = {1316},
  month = apr,
  year = {2024}
}

@article{Gluza26,
  title = {Double-Bracket Quantum Algorithms for Quantum Imaginary-Time Evolution},
  author = {Gluza, Marek and Son, Jeongrak and Tiang, Bi Hong and Zander, Ren\'e and Seidel, Raphael and Suzuki, Yudai and Holmes, Zo\"e and Ng, Nelly H. Y.},
  journal = {Phys. Rev. Lett.},
  volume = {136},
  issue = {2},
  pages = {020601},
  numpages = {12},
  year = {2026},
  month = {Jan},
  publisher = {American Physical Society},
  doi = {10.1103/rw81-k8vk},
  url = {https://link.aps.org/doi/10.1103/rw81-k8vk}
}

@article{Brody02,
    author = {Brody, Dorje C. and Hughston, Lane P.},
    title = {Efficient simulation of quantum state reduction},
    journal = {Journal of Mathematical Physics},
    volume = {43},
    number = {11},
    pages = {5254-5261},
    year = {2002},
    month = {11},
    issn = {0022-2488},
    doi = {10.1063/1.1512975},
    url = {https://doi.org/10.1063/1.1512975}
}

@article{Adler01,
doi = {10.1088/0305-4470/34/42/306},
url = {https://doi.org/10.1088/0305-4470/34/42/306},
year = {2001},
month = {oct},
publisher = {},
volume = {34},
number = {42},
pages = {8795},
author = {S. L. Adler and D. C. Brody and T. A. Brun and L. P. Hughston},
title = {Martingale
models for quantum state reduction},
journal = {Journal of Physics A: Mathematical and General},
}

@article{Brody06,
doi = {10.1088/0305-4470/39/4/008},
url = {https://doi.org/10.1088/0305-4470/39/4/008},
year = {2006},
month = {jan},
publisher = {},
volume = {39},
number = {4},
pages = {833},
author = {Brody, Dorje C and Hughston, Lane P},
title = {Quantum noise and stochastic reduction},
journal = {Journal of Physics A: Mathematical and General},
}

@article{Hillery84,
title = {Distribution functions in physics: Fundamentals},
journal = {Physics Reports},
volume = {106},
number = {3},
pages = {121-167},
year = {1984},
issn = {0370-1573},
doi = {https://doi.org/10.1016/0370-1573(84)90160-1},
url = {https://www.sciencedirect.com/science/article/pii/0370157384901601},
author = {M. Hillery and R.F. O'Connell and M.O. Scully and E.P. Wigner},
}

@article{MatsoukasRoubeas2024,
  doi = {10.22331/q-2024-08-27-1446},
  url = {https://doi.org/10.22331/q-2024-08-27-1446},
  title = {Quantum {C}haos and {C}oherence: {R}andom {P}arametric {Q}uantum {C}hannels},
  author = {Matsoukas-Roubeas, Apollonas S. and Prosen, Toma{\v{z}} and del Campo, Adolfo },
  journal = {{Quantum}},
  issn = {2521-327X},
  publisher = {{Verein zur F{\"{o}}rderung des Open Access Publizierens in den Quantenwissenschaften}},
  volume = {8},
  pages = {1446},
  month = aug,
  year = {2024}
}

@article{delCampo2019,
doi = {10.1088/1367-2630/ab1437},
url = {https://doi.org/10.1088/1367-2630/ab1437},
year = {2019},
month = {may},
publisher = {IOP Publishing},
volume = {21},
number = {5},
pages = {050201},
author = {del Campo, Adolfo and Kim, Kihwan},
title = {Focus on Shortcuts to Adiabaticity},
journal = {New Journal of Physics},
}

@article{Kiely2021,
doi = {10.1209/0295-5075/134/10001},
url = {https://doi.org/10.1209/0295-5075/134/10001},
year = {2021},
month = {may},
publisher = {EDP Sciences, IOP Publishing and Società Italiana di Fisica},
volume = {134},
number = {1},
pages = {10001},
author = {Kiely, Anthony},
title = {Exact classical noise master equations: Applications and connections},
journal = {Europhysics Letters},
}

@article{Ai2021,
  title = {Experimental verification of anti--{K}ibble-{Z}urek behavior in a quantum system under a noisy control field},
  author = {Ai, Ming-Zhong and Cui, Jin-Ming and He, Ran and Qian, Zhong-Hua and Gao, Xin-Xia and Huang, Yun-Feng and Li, Chuan-Feng and Guo, Guang-Can},
  journal = {Phys. Rev. A},
  volume = {103},
  issue = {1},
  pages = {012608},
  numpages = {6},
  year = {2021},
  month = {Jan},
  publisher = {American Physical Society},
  doi = {10.1103/PhysRevA.103.012608},
  url = {https://link.aps.org/doi/10.1103/PhysRevA.103.012608}
}

@article{Iwamura24,
  title = {Analytical derivation and extension of the anti-{K}ibble-{Z}urek scaling in the transverse field {I}sing model},
  author = {Iwamura, Kaito and Suzuki, Takayuki},
  journal = {Phys. Rev. B},
  volume = {110},
  issue = {14},
  pages = {144102},
  numpages = {16},
  year = {2024},
  month = {Oct},
  publisher = {American Physical Society},
  doi = {10.1103/PhysRevB.110.144102},
  url = {https://link.aps.org/doi/10.1103/PhysRevB.110.144102}
}

@misc{Xu25,
      title={Quantum-to-Classical Transition via Single-Shot Generalized Measurements}, 
      author={Zhenyu Xu},
      year={2025},
      eprint={2507.13174},
      archivePrefix={arXiv},
      primaryClass={quant-ph},
      url={https://arxiv.org/abs/2507.13174}, 
}

@misc{villanueva2025,
      title={Hamiltonian and double-bracket flow formulations of quantum measurements}, 
      author={Aarón Villanueva and Luis Pedro García-Pintos},
      year={2025},
      eprint={2512.15412},
      archivePrefix={arXiv},
      primaryClass={quant-ph},
      url={https://arxiv.org/abs/2512.15412}, 
}

@Article{delCampo2020,
author={del Campo, Adolfo and {Takayanagi}, Tadashi},
title={{Decoherence} in {Conformal} {Field} {Theory}},
journal={JHEP},
year={2020},
month={Feb},
day={26},
volume={2020},
number={2},
pages={170},
issn={1029-8479},
doi={10.1007/JHEP02(2020)170},
url={https://doi.org/10.1007/JHEP02(2020)170}
}

@article{Cabrera15,
  title = {Efficient method to generate time evolution of the {W}igner function for open quantum systems},
  author = {Cabrera, Renan and Bondar, Denys I. and Jacobs, Kurt and Rabitz, Herschel A.},
  journal = {Phys. Rev. A},
  volume = {92},
  issue = {4},
  pages = {042122},
  numpages = {10},
  year = {2015},
  month = {Oct},
  publisher = {American Physical Society},
  doi = {10.1103/PhysRevA.92.042122},
  url = {https://link.aps.org/doi/10.1103/PhysRevA.92.042122}
}

@article{HuPazZhang92,
  title = {Quantum Brownian motion in a general environment: Exact master equation with nonlocal dissipation and colored noise},
  author = {Hu, B. L. and Paz, Juan Pablo and Zhang, Yuhong},
  journal = {Phys. Rev. D},
  volume = {45},
  issue = {8},
  pages = {2843--2861},
  numpages = {0},
  year = {1992},
  month = {Apr},
  publisher = {American Physical Society},
  doi = {10.1103/PhysRevD.45.2843},
  url = {https://link.aps.org/doi/10.1103/PhysRevD.45.2843}
}

@article{Isar96,
author = {Isar, A. and Sandulescu, A. and Scheid, W.},
title = {PHASE SPACE REPRESENTATION FOR OPEN QUANTUM SYSTEMS WITHIN THE {L}INDBLAD THEORY},
journal = {International Journal of Modern Physics B},
volume = {10},
number = {22},
pages = {2767-2779},
year = {1996},
doi = {10.1142/S0217979296001240},
URL = {https://doi.org/10.1142/S0217979296001240}
}

@article{Brody25,
  title = {Phase-Space Measurements, Decoherence, and Classicality},
  author = {Brody, Dorje C. and Graefe, Eva-Maria and Melanathuru, Rishindra},
  journal = {Phys. Rev. Lett.},
  volume = {134},
  issue = {12},
  pages = {120201},
  numpages = {6},
  year = {2025},
  month = {Mar},
  publisher = {American Physical Society},
  doi = {10.1103/PhysRevLett.134.120201},
  url = {https://link.aps.org/doi/10.1103/PhysRevLett.134.120201}
}

@article{Zurek91,
  author       = {Zurek, W. H.},
  title        = {Decoherence and the Transition from Quantum to Classical},
  journal      = {Physics Today},
  volume       = {44},
  number       = {10},
  pages        = {36--44},
  year         = {1991},
  doi          = {10.1063/1.881293},
  url          = {https://doi.org/10.1063/1.881293}
}

@article{Lidar98,
  title = {Decoherence-Free Subspaces for Quantum Computation},
  author = {Lidar, D. A. and Chuang, I. L. and Whaley, K. B.},
  journal = {Phys. Rev. Lett.},
  volume = {81},
  issue = {12},
  pages = {2594--2597},
  numpages = {0},
  year = {1998},
  month = {Sep},
  publisher = {American Physical Society},
  doi = {10.1103/PhysRevLett.81.2594},
  url = {https://link.aps.org/doi/10.1103/PhysRevLett.81.2594}
}

@article{Beau17,
  title = {Nonexponential Quantum Decay under Environmental Decoherence},
  author = {Beau, M. and Kiukas, J. and Egusquiza, I. L. and del Campo, A.},
  journal = {Phys. Rev. Lett.},
  volume = {119},
  issue = {13},
  pages = {130401},
  numpages = {6},
  year = {2017},
  month = {Sep},
  publisher = {American Physical Society},
  doi = {10.1103/PhysRevLett.119.130401},
  url = {https://link.aps.org/doi/10.1103/PhysRevLett.119.130401}
}

@article{Beau17metro,
  title = {Nonlinear Quantum Metrology of Many-Body Open Systems},
  author = {Beau, M. and del Campo, A.},
  journal = {Phys. Rev. Lett.},
  volume = {119},
  issue = {1},
  pages = {010403},
  numpages = {6},
  year = {2017},
  month = {Jul},
  publisher = {American Physical Society},
  doi = {10.1103/PhysRevLett.119.010403},
  url = {https://link.aps.org/doi/10.1103/PhysRevLett.119.010403}
}

@article{YangXu24,
  title = {Decoherence rate in random {L}indblad dynamics},
  author = {Yang, Yifeng and Xu, Zhenyu and del Campo, Adolfo},
  journal = {Phys. Rev. Res.},
  volume = {6},
  issue = {2},
  pages = {023229},
  numpages = {11},
  year = {2024},
  month = {Jun},
  publisher = {American Physical Society},
  doi = {10.1103/PhysRevResearch.6.023229},
  url = {https://link.aps.org/doi/10.1103/PhysRevResearch.6.023229}
}

@article{korbicz2017generic,
  title = {Generic appearance of objective results in quantum measurements},
  author = {Korbicz, J. K. and Aguilar, E. A. and \ifmmode \acute{C}\else \'{C}\fi{}wikli\ifmmode \acute{n}\else \'{n}\fi{}ski, P. and Horodecki, P.},
  journal = {Phys. Rev. A},
  volume = {96},
  issue = {3},
  pages = {032124},
  numpages = {11},
  year = {2017},
  month = {Sep},
  publisher = {American Physical Society},
  doi = {10.1103/PhysRevA.96.032124},
  url = {https://link.aps.org/doi/10.1103/PhysRevA.96.032124}
}

@article{apollo2023unitarity,
  title = {Unitarity breaking in self-averaging spectral form factors},
  author = {Matsoukas-Roubeas, Apollonas S. and Beau, Mathieu and Santos, Lea F. and del Campo, Adolfo},
  journal = {Phys. Rev. A},
  volume = {108},
  issue = {6},
  pages = {062201},
  year = {2023},
  month = {Dec},
  publisher = {American Physical Society},
  doi = {10.1103/PhysRevA.108.062201},
  url = {https://link.aps.org/doi/10.1103/PhysRevA.108.062201}
}

@article{gisin1984quantum,
  title = {Quantum Measurements and Stochastic Processes},
  author = {Gisin, N.},
  journal = {Phys. Rev. Lett.},
  volume = {52},
  issue = {19},
  pages = {1657--1660},
  numpages = {0},
  year = {1984},
  month = {May},
  publisher = {American Physical Society},
  doi = {10.1103/PhysRevLett.52.1657},
  url = {https://link.aps.org/doi/10.1103/PhysRevLett.52.1657}
}

@article{schneider1998decoherence,
  title = {Decoherence in ion traps due to laser intensity and phase fluctuations},
  author = {Schneider, S. and Milburn, G. J.},
  journal = {Phys. Rev. A},
  volume = {57},
  issue = {5},
  pages = {3748--3752},
  numpages = {0},
  year = {1998},
  month = {May},
  publisher = {American Physical Society},
  doi = {10.1103/PhysRevA.57.3748},
  url = {https://link.aps.org/doi/10.1103/PhysRevA.57.3748}
}

@article{adler2003weisskopf,
  title = {Weisskopf-{W}igner decay theory for the energy-driven stochastic {S}chr\"odinger equation},
  author = {Adler, Stephen L.},
  journal = {Phys. Rev. D},
  volume = {67},
  issue = {2},
  pages = {025007},
  numpages = {14},
  year = {2003},
  month = {Jan},
  publisher = {American Physical Society},
  doi = {10.1103/PhysRevD.67.025007},
  url = {https://link.aps.org/doi/10.1103/PhysRevD.67.025007}
}

@article{Dutta16,
  title = {Anti-{K}ibble-{Z}urek Behavior in Crossing the Quantum Critical Point of a Thermally Isolated System Driven by a Noisy Control Field},
  author = {Dutta, Anirban and Rahmani, Armin and del Campo, Adolfo},
  journal = {Phys. Rev. Lett.},
  volume = {117},
  issue = {8},
  pages = {080402},
  numpages = {5},
  year = {2016},
  month = {Aug},
  publisher = {American Physical Society},
  doi = {10.1103/PhysRevLett.117.080402},
  url = {https://link.aps.org/doi/10.1103/PhysRevLett.117.080402}
}

@ARTICLE{Budini01,
  author = {Budini, Adri\'an A.},
  title = {Quantum systems subject to the action of classical stochastic fields},
  journal = {Phys. Rev. A},
  year = {2001},
  volume = {64},
  pages = {052110},
  month = {Oct},
  doi = {10.1103/PhysRevA.64.052110},
  issue = {5},
  numpages = {12},
  publisher = {American Physical Society},
  url = {https://link.aps.org/doi/10.1103/PhysRevA.64.052110}
}

@article{Agarwal69,
  title = {Master Equations in Phase-Space Formulation of Quantum Optics},
  author = {Agarwal, G. S.},
  journal = {Phys. Rev.},
  volume = {178},
  issue = {5},
  pages = {2025--2035},
  numpages = {0},
  year = {1969},
  month = {Feb},
  publisher = {American Physical Society},
  doi = {10.1103/PhysRev.178.2025},
  url = {https://link.aps.org/doi/10.1103/PhysRev.178.2025}
}

@article{Dekker77,
  title = {Quantization of the linearly damped harmonic oscillator},
  author = {Dekker, H.},
  journal = {Phys. Rev. A},
  volume = {16},
  issue = {5},
  pages = {2126--2134},
  numpages = {0},
  year = {1977},
  month = {Nov},
  publisher = {American Physical Society},
  doi = {10.1103/PhysRevA.16.2126},
  url = {https://link.aps.org/doi/10.1103/PhysRevA.16.2126}
}

@article{Caldeira83,
title = {Path integral approach to quantum Brownian motion},
journal = {Physica A: Statistical Mechanics and its Applications},
volume = {121},
number = {3},
pages = {587-616},
year = {1983},
issn = {0378-4371},
doi = {https://doi.org/10.1016/0378-4371(83)90013-4},
url = {https://www.sciencedirect.com/science/article/pii/0378437183900134},
author = {A.O. Caldeira and A.J. Leggett},
}

@article{Unruh89,
  title = {Reduction of a wave packet in quantum Brownian motion},
  author = {Unruh, W. G. and Zurek, W. H.},
  journal = {Phys. Rev. D},
  volume = {40},
  issue = {4},
  pages = {1071--1094},
  numpages = {0},
  year = {1989},
  month = {Aug},
  publisher = {American Physical Society},
  doi = {10.1103/PhysRevD.40.1071},
  url = {https://link.aps.org/doi/10.1103/PhysRevD.40.1071}
}

@article{Steuernagel15,
    author = {Steuernagel, Ole and Lee, Ray-Kuang},
    title = {Lindblad superoperators from {W}igner’s phase space continuity equation},
    journal = {AIP Advances},
    volume = {15},
    number = {1},
    pages = {015009},
    year = {2025},
    month = {01},
    issn = {2158-3226},
    doi = {10.1063/5.0244814},
    url = {https://doi.org/10.1063/5.0244814}
}

@article{Braasch19,
  title = {Wigner current for open quantum systems},
  author = {Braasch, William F. and Friedman, Oscar D. and Rimberg, Alexander J. and Blencowe, Miles P.},
  journal = {Phys. Rev. A},
  volume = {100},
  issue = {1},
  pages = {012124},
  numpages = {10},
  year = {2019},
  month = {Jul},
  publisher = {American Physical Society},
  doi = {10.1103/PhysRevA.100.012124},
  url = {https://link.aps.org/doi/10.1103/PhysRevA.100.012124}
}

@Article{Bondar2016,
author={Bondar, Denys I.
and Cabrera, Renan
and Campos, Andre
and Mukamel, Shaul
and Rabitz, Herschel A.},
title={Wigner--{L}indblad Equations for Quantum Friction},
journal={The Journal of Physical Chemistry Letters},
year={2016},
month={May},
day={05},
publisher={American Chemical Society},
volume={7},
number={9},
pages={1632-1637},
doi={10.1021/acs.jpclett.6b00498},
url={https://doi.org/10.1021/acs.jpclett.6b00498}
}

@article{Jacobs2006,
   title={A straightforward introduction to continuous quantum measurement},
   volume={47},
   ISSN={1366-5812},
   url={http://dx.doi.org/10.1080/00107510601101934},
   DOI={10.1080/00107510601101934},
   number={5},
   journal={Contemp. Phys.},
   publisher={Informa UK Limited},
   author={Jacobs, Kurt and Steck, Daniel A.},
   year={2006},
   month={Sep},
   pages={279–303}
}

@book{Jacobs2014, 
    place={Cambridge}, 
    title={Quantum Measurement Theory and its Applications}, publisher={Cambridge University Press}, 
    author={Jacobs, Kurt}, 
    doi={https://doi.org/10.1017/CBO9781139179027},
    year={2014}
}

@incollection{rossmann2023,
    author = {Rossmann, Wulf},
    isbn = {9780198596837},
    booktitle = {Lie Groups: An Introduction Through Linear Groups},
    publisher = {Oxford University Press},
    year = {2002},
    month = {01},
    doi = {10.1093/oso/9780198596837.002.0001},
    url = {https://doi.org/10.1093/oso/9780198596837.002.0001}}

@book{Helmke_Moore_2014, 
place={London}, 
title={Optimization and dynamical systems}, 
publisher={Springer}, 
author={Helmke, Uwe and Moore, John B.}, 
year={2014},
doi={https://doi.org/10.1007/978-1-4471-3467-1}}

@book{Bloch_1994, 
place={Providence, RI}, 
title={Hamiltonian and gradient flows, algorithms and control}, 
publisher={American Mathematical Society}, 
author={Bloch, Anthony M.}, 
year={1994},
doi={https://doi.org/10.1090/fic/003}}

@article{helmkereview,
author = {Schulte-Herbr\"{u}ggen, Thomas and Glaser, Steffen J. and Dirr, Gunther and Helmke, Uwe},
title = {Gradient flows for optimization in quantum information and quantum dynamics: foundations and applications},
journal = {Reviews in Mathematical Physics},
volume = {22},
number = {06},
pages = {597-667},
year = {2010},
doi = {10.1142/S0129055X10004053},
URL = { https://doi.org/10.1142/S0129055X10004053}}

@ARTICLE{Budini00,
  author = {Budini, Adri\'an A.},
  title = {Non-{M}arkovian {G}aussian dissipative stochastic wave vector},
  journal = {Phys. Rev. A},
  year = {2000},
  volume = {63},
  pages = {012106},
  month = {Dec},
  doi = {10.1103/PhysRevA.63.012106},
  issue = {1},
  numpages = {11},
  publisher = {American Physical Society},
  url = {https://link.aps.org/doi/10.1103/PhysRevA.63.012106}
}

@article{Ruschhaupt2012,
doi = {10.1088/1367-2630/14/9/093040},
url = {https://doi.org/10.1088/1367-2630/14/9/093040},
year = {2012},
month = {sep},
publisher = {IOP Publishing},
volume = {14},
number = {9},
pages = {093040},
author = {Ruschhaupt, A and Chen, Xi and Alonso, D and Muga, J G},
title = {Optimally robust shortcuts to population inversion in two-level quantum systems},
journal = {New Journal of Physics},
}

@book{VanKampen92,
	author		= {N. G. van Kampen},
	title		= {Stochastic processes in physics and chemistry},
	publisher	= {Elsevier Science B. V.},
	year		= {1992},
        doi={10.1016/B978-0-444-52965-7.X5000-4},
        url={https://doi.org/10.1016/B978-0-444-52965-7.X5000-4},
	address		= {Amsterdam, The Netherlands}
}

@article{GKS76,
    author = {Gorini, Vittorio and Kossakowski, Andrzej and Sudarshan, E. C. G.},
    title = {Completely positive dynamical semigroups of N‐level systems},
    journal = {Journal of Mathematical Physics},
    volume = {17},
    number = {5},
    pages = {821-825},
    year = {1976},
    month = {05},
    issn = {0022-2488},
    doi = {10.1063/1.522979},
    url = {https://doi.org/10.1063/1.522979}
}

@article{Wang25a,
doi = {10.1088/1751-8121/adc2bb},
url = {https://doi.org/10.1088/1751-8121/adc2bb},
year = {2025},
month = {mar},
publisher = {IOP Publishing},
volume = {58},
number = {13},
pages = {135303},
author = {Wang, Pei},
title = {Random non-{H}ermitian action theory for stochastic quantum dynamics: from canonical to path integral quantization},
journal = {Journal of Physics A: Mathematical and Theoretical},
}

@article{Wang25b,
  title = {Random non-{H}ermitian {H}amiltonian framework for symmetry-breaking dynamics},
  author = {Wang, Pei},
  journal = {Phys. Rev. A},
  volume = {112},
  issue = {1},
  pages = {012209},
  numpages = {15},
  year = {2025},
  month = {Jul},
  publisher = {American Physical Society},
  doi = {10.1103/rkdn-j6hy},
  url = {https://link.aps.org/doi/10.1103/rkdn-j6hy}
}

@article{Pearle89,
  title = {Combining stochastic dynamical state-vector reduction with spontaneous localization},
  author = {Pearle, Philip},
  journal = {Phys. Rev. A},
  volume = {39},
  issue = {5},
  pages = {2277--2289},
  numpages = {0},
  year = {1989},
  month = {Mar},
  publisher = {American Physical Society},
  doi = {10.1103/PhysRevA.39.2277},
  url = {https://link.aps.org/doi/10.1103/PhysRevA.39.2277}
}

@article{bassi2003dynamical,
title = {Dynamical reduction models},
journal = {Physics Reports},
volume = {379},
number = {5},
pages = {257-426},
year = {2003},
issn = {0370-1573},
doi = {https://doi.org/10.1016/S0370-1573(03)00103-0},
url = {https://www.sciencedirect.com/science/article/pii/S0370157303001030},
author = {Angelo Bassi and GianCarlo Ghirardi}
}

@article{bassi2013models,
  title = {Models of wave-function collapse, underlying theories, and experimental tests},
  author = {Bassi, Angelo and Lochan, Kinjalk and Satin, Seema and Singh, Tejinder P. and Ulbricht, Hendrik},
  journal = {Rev. Mod. Phys.},
  volume = {85},
  issue = {2},
  pages = {471--527},
  numpages = {0},
  year = {2013},
  month = {Apr},
  publisher = {American Physical Society},
  doi = {10.1103/RevModPhys.85.471},
  url = {https://link.aps.org/doi/10.1103/RevModPhys.85.471}
}

@Article{MatsoukasRoubeas23,
author={Matsoukas-Roubeas, Apollonas S. and Roccati, Federico and Cornelius, Julien and Xu, Zhenyu and Chenu, Aur{\'e}lia and del Campo, Adolfo},
title={Non-{H}ermitian {H}amiltonian deformations in quantum mechanics},
journal={JHEP},
year={2023},
month={Jan},
day={13},
volume={2023},
number={1},
pages={60},
issn={1029-8479},
doi={10.1007/JHEP01(2023)060},
url={https://doi.org/10.1007/JHEP01(2023)060}
}

@article{Xu21SFF,
  title = {Thermofield dynamics: Quantum chaos versus decoherence},
  author = {Xu, Z. and Chenu, A. and Prosen, T. and del Campo, A.},
  journal = {Phys. Rev. B},
  volume = {103},
  issue = {6},
  pages = {064309},
  numpages = {11},
  year = {2021},
  month = {Feb},
  publisher = {American Physical Society},
  doi = {10.1103/PhysRevB.103.064309},
  url = {https://link.aps.org/doi/10.1103/PhysRevB.103.064309}
}

@article{Cornelius21,
  title = {Spectral Filtering Induced by Non-{H}ermitian Evolution with Balanced Gain and Loss: Enhancing Quantum Chaos},
  author = {Cornelius, Julien and Xu, Zhenyu and Saxena, Avadh and Chenu, Aur\'elia and del Campo, Adolfo},
  journal = {Phys. Rev. Lett.},
  volume = {128},
  issue = {19},
  pages = {190402},
  numpages = {6},
  year = {2022},
  month = {May},
  publisher = {American Physical Society},
  doi = {10.1103/PhysRevLett.128.190402},
  url = {https://link.aps.org/doi/10.1103/PhysRevLett.128.190402}
}

@article{percival1994primary,
author = {Percival, Ian Colin },
title = {Primary state diffusion},
journal = {Proceedings of the Royal Society of London. Series A: Mathematical and Physical Sciences},
volume = {447},
number = {1929},
pages = {189-209},
year = {1994},
doi = {10.1098/rspa.1994.0135},
URL = {https://royalsocietypublishing.org/doi/abs/10.1098/rspa.1994.0135},
}

@article{Brockett91,
title = {Dynamical systems that sort lists, diagonalize matrices, and solve linear programming problems},
journal = {Linear Algebra and its Applications},
volume = {146},
pages = {79-91},
year = {1991},
issn = {0024-3795},
doi = {https://doi.org/10.1016/0024-3795(91)90021-N},
url = {https://www.sciencedirect.com/science/article/pii/002437959190021N},
author = {R. W. Brockett},
}

@article{Hornedal23,
  doi = {10.22331/q-2023-07-11-1055},
  url = {https://doi.org/10.22331/q-2023-07-11-1055},
  title = {Geometric {O}perator {Q}uantum {S}peed {L}imit, {W}egner {H}amiltonian {F}low and {O}perator {G}rowth},
  author = {H{\"{o}}rnedal, Niklas and Carabba, Nicoletta and Takahashi, Kazutaka and del Campo, Adolfo},
  journal = {{Quantum}},
  issn = {2521-327X},
  publisher = {{Verein zur F{\"{o}}rderung des Open Access Publizierens in den Quantenwissenschaften}},
  volume = {7},
  pages = {1055},
  month = jul,
  year = {2023}
}

@article{Wegner94,
author = {Wegner, F.},
title = {Flow-equations for {H}amiltonians},
journal = {Annalen der Physik},
volume = {506},
number = {2},
pages = {77-91},
doi = {https://doi.org/10.1002/andp.19945060203},
year = {1994}
}

@article{Wegner2001,
	author = {F. J. Wegner},
	doi = {https://doi.org/10.1016/S0370-1573(00)00136-8},
	issn = {0370-1573},
	journal = {Physics Reports},
	number = {1},
	pages = {77-89},
	title = {Flow equations for {H}amiltonians},
	url = {https://www.sciencedirect.com/science/article/pii/S0370157300001368},
	volume = {348},
	year = {2001},}

@book{Kehrein2007,
  title={The Flow Equation Approach to Many-Particle Systems},
  author={Kehrein, S.},
  isbn={9783540340683},
  lccn={2006925894},
  series={Springer Tracts in Modern Physics},
  url={https://link.springer.com/book/10.1007/3-540-34068-8},
  year={2007},
  publisher={Springer Berlin Heidelberg}
}

@book{Zachos05,
  title        = {Quantum Mechanics in Phase Space: An Overview with Selected Papers},
  editor       = {Zachos, Cosmas K. and Fairlie, David B. and Curtright, Thomas L.},
  publisher    = {World Scientific},
  address      = {Singapore},
  year         = {2005},
  doi          = {10.1142/5287},
  isbn         = {978-981-238-384-6},
  url          = {https://www.worldscientific.com/worldscibooks/10.1142/5287}
}

@article{Wigner32,
  title = {On the Quantum Correction For Thermodynamic Equilibrium},
  author = {Wigner, E.},
  journal = {Phys. Rev.},
  volume = {40},
  issue = {5},
  pages = {749--759},
  numpages = {0},
  year = {1932},
  month = {Jun},
  publisher = {American Physical Society},
  doi = {10.1103/PhysRev.40.749},
  url = {https://link.aps.org/doi/10.1103/PhysRev.40.749}
}

@article{GlazekWilson93,
  title = {Renormalization of {H}amiltonians},
  author = {G\l{}azek, S. D. and Wilson, K. G.},
  journal = {Phys. Rev. D},
  volume = {48},
  issue = {12},
  pages = {5863--5872},
  numpages = {0},
  year = {1993},
  month = {Dec},
  publisher = {American Physical Society},
  doi = {10.1103/PhysRevD.48.5863},
  url = {https://link.aps.org/doi/10.1103/PhysRevD.48.5863}
}

@article{GlazekWilson94,
  title = {Perturbative renormalization group for {H}amiltonians},
  author = {Glazek, S. D. and Wilson, K. G.},
  journal = {Phys. Rev. D},
  volume = {49},
  issue = {8},
  pages = {4214--4218},
  numpages = {0},
  year = {1994},
  month = {Apr},
  publisher = {American Physical Society},
  doi = {10.1103/PhysRevD.49.4214},
  url = {https://link.aps.org/doi/10.1103/PhysRevD.49.4214}
}

@article{Moyal1949quantum, 
title={Quantum mechanics as a statistical theory}, 
volume={45}, 
DOI={10.1017/S0305004100000487}, 
number={1}, 
journal={Mathematical Proceedings of the Cambridge Philosophical Society}, 
author={Moyal, J. E.}, 
year={1949}, 
pages={99–124}}

@article{Zurek:1994wd,
title = {Decoherence, chaos, and the second law},
  author = {Zurek, Wojciech Hubert and Paz, Juan Pablo},
  journal = {Phys. Rev. Lett.},
  volume = {72},
  issue = {16},
  pages = {2508--2511},
  numpages = {0},
  year = {1994},
  month = {Apr},
  publisher = {American Physical Society},
  doi = {10.1103/PhysRevLett.72.2508},
  url = {https://link.aps.org/doi/10.1103/PhysRevLett.72.2508%7D
}}

@article{jensen1990wigner,
  title = {Wigner symbols, quantum dynamics, and the kicked rotator},
  author = {Jensen, J. H. and Niu, Q.},
  journal = {Phys. Rev. A},
  volume = {42},
  issue = {5},
  pages = {2513--2519},
  numpages = {0},
  year = {1990},
  month = {Sep},
  publisher = {American Physical Society},
  doi = {10.1103/PhysRevA.42.2513},
  url = {https://link.aps.org/doi/10.1103/PhysRevA.42.2513},
}

@article{Zurek03,
  title = {Decoherence, einselection, and the quantum origins of the classical},
  author = {Zurek, Wojciech Hubert},
  journal = {Rev. Mod. Phys.},
  volume = {75},
  issue = {3},
  pages = {715--775},
  numpages = {0},
  year = {2003},
  month = {May},
  publisher = {American Physical Society},
  doi = {10.1103/RevModPhys.75.715},
  url = {https://link.aps.org/doi/10.1103/RevModPhys.75.715}
}

@article{kolovsky1996quantum,
    author = {Kolovsky, Andrey R.},
    title = {Quantum coherence, evolution of the {W}igner function, and transition from quantum to classical dynamics for a chaotic system},
    journal = {Chaos: An Interdisciplinary Journal of Nonlinear Science},
    volume = {6},
    number = {4},
    pages = {534-542},
    year = {1996},
    month = {12},
    issn = {1054-1500},
    doi = {10.1063/1.166201},
    url = {https://doi.org/10.1063/1.166201}
}

@article{Habib:1998ai,
  title = {Decoherence, Chaos, and the Correspondence Principle},
  author = {Habib, Salman and Shizume, Kosuke and Zurek, Wojciech Hubert},
  journal = {Phys. Rev. Lett.},
  volume = {80},
  issue = {20},
  pages = {4361--4365},
  numpages = {0},
  year = {1998},
  month = {May},
  publisher = {American Physical Society},
  doi = {10.1103/PhysRevLett.80.4361},
  url = {https://link.aps.org/doi/10.1103/PhysRevLett.80.4361}
}

@article{lutkenhaus1995nonclassical,
  title={Nonclassical effects in phase space},
  author={L{\"u}tkenhaus, N and Barnett, Stephen M},
  journal={Physical Review A},
  volume={51},
  number={4},
  pages={3340},
  year={1995},
  publisher={APS},
url = {https://doi.org/10.1103/PhysRevA.51.3340}
}

@article{strandberg2019numerical,
  title = {Numerical study of {W}igner negativity in one-dimensional steady-state resonance fluorescence},
  author = {Strandberg, Ingrid and Lu, Yong and Quijandr\'{\i}a, Fernando and Johansson, G\"oran},
  journal = {Phys. Rev. A},
  volume = {100},
  issue = {6},
  pages = {063808},
  numpages = {15},
  year = {2019},
  month = {Dec},
  publisher = {American Physical Society},
  doi = {10.1103/PhysRevA.100.063808},
  url = {https://link.aps.org/doi/10.1103/PhysRevA.100.063808}
}

@article{pashayan2015estimating,
  title = {Estimating Outcome Probabilities of Quantum Circuits Using Quasiprobabilities},
  author = {Pashayan, Hakop and Wallman, Joel J. and Bartlett, Stephen D.},
  journal = {Phys. Rev. Lett.},
  volume = {115},
  issue = {7},
  pages = {070501},
  numpages = {5},
  year = {2015},
  month = {Aug},
  publisher = {American Physical Society},
  doi = {10.1103/PhysRevLett.115.070501},
  url = {https://link.aps.org/doi/10.1103/PhysRevLett.115.070501}
}

@article{albarelli2018resource,
  title = {Resource theory of quantum non-{G}aussianity and {W}igner negativity},
  author = {Albarelli, Francesco and Genoni, Marco G. and Paris, Matteo G. A. and Ferraro, Alessandro},
  journal = {Phys. Rev. A},
  volume = {98},
  issue = {5},
  pages = {052350},
  numpages = {17},
  year = {2018},
  month = {Nov},
  publisher = {American Physical Society},
  doi = {10.1103/PhysRevA.98.052350},
  url = {https://link.aps.org/doi/10.1103/PhysRevA.98.052350}
}

@article{smithey1993measurement,
  title = {Measurement of the {W}igner distribution and the density matrix of a light mode using optical homodyne tomography: Application to squeezed states and the vacuum},
  author = {Smithey, D. T. and Beck, M. and Raymer, M. G. and Faridani, A.},
  journal = {Phys. Rev. Lett.},
  volume = {70},
  issue = {9},
  pages = {1244--1247},
  numpages = {0},
  year = {1993},
  month = {Mar},
  publisher = {American Physical Society},
  doi = {10.1103/PhysRevLett.70.1244},
  url = {https://link.aps.org/doi/10.1103/PhysRevLett.70.1244}
}

@article{nogues2000measurement,
  title = {Measurement of a negative value for the {W}igner function of radiation},
  author = {Nogues, G. and Rauschenbeutel, A. and Osnaghi, S. and Bertet, P. and Brune, M. and Raimond, J. M. and Haroche, S. and Lutterbach, L. G. and Davidovich, L.},
  journal = {Phys. Rev. A},
  volume = {62},
  issue = {5},
  pages = {054101},
  numpages = {4},
  year = {2000},
  month = {Oct},
  publisher = {American Physical Society},
  doi = {10.1103/PhysRevA.62.054101},
  url = {https://link.aps.org/doi/10.1103/PhysRevA.62.054101}
}

@article{Kenfack:2004ges,
doi = {10.1088/1464-4266/6/10/003},
url = {https://doi.org/10.1088/1464-4266/6/10/003},
year = {2004},
month = {aug},
publisher = {},
volume = {6},
number = {10},
pages = {396},
author = {Anatole Kenfack and Karol Życzkowski},
title = {Negativity of the {W}igner function as an indicator of non-classicality},
journal = {Journal of Optics B: Quantum and Semiclassical Optics},
}

@article{PhysRevLett.65.2927,
  title = {Quantum tunneling and chaos in a driven anharmonic oscillator},
  author = {Lin, W. A. and Ballentine, L. E.},
  journal = {Phys. Rev. Lett.},
  volume = {65},
  issue = {24},
  pages = {2927--2930},
  numpages = {0},
  year = {1990},
  month = {Dec},
  publisher = {American Physical Society},
  doi = {10.1103/PhysRevLett.65.2927},
  url = {https://link.aps.org/doi/10.1103/PhysRevLett.65.2927}
}

@article{gerry1997quantum,
  title={Quantum superpositions and Schr{\"o}dinger cat states in quantum optics},
  author={Gerry, CC and Knight, PL},
  journal={American Journal of Physics},
  volume={65},
  number={10},
  pages={964--974},
  year={1997},
  publisher={American Association of Physics Teachers},
doi = {https://doi.org/10.1119/1.18698}
}

@article{Wiersma2023optimizing,
  title = {Optimizing quantum circuits with {R}iemannian gradient flow},
  author = {Wiersema, Roeland and Killoran, Nathan},
  journal = {Phys. Rev. A},
  volume = {107},
  issue = {6},
  pages = {062421},
  numpages = {11},
  year = {2023},
  month = {Jun},
  publisher = {American Physical Society},
  doi = {10.1103/PhysRevA.107.062421},
  url = {https://link.aps.org/doi/10.1103/PhysRevA.107.062421}
}

@article{Wiersema2024here,
  doi = {10.22331/q-2024-03-07-1275},
  url = {https://doi.org/10.22331/q-2024-03-07-1275},
  title = {Here comes the {SU}({N}): multivariate quantum gates and gradients},
  author = {Wiersema, Roeland and Lewis, Dylan and Wierichs, David and Carrasquilla, Juan and Killoran, Nathan},
  journal = {{Quantum}},
  issn = {2521-327X},
  publisher = {{Verein zur F{\"{o}}rderung des Open Access Publizierens in den Quantenwissenschaften}},
  volume = {8},
  pages = {1275},
  month = mar,
  year = {2024}
}

@article{Kaplanek2025Gradient,
  title = {Lindblad evolution as gradient flow},
  author = {Kaplanek, Greg and Maloney, Alexander and Pollack, Jason and VanAllen, Dylan},
  journal = {Phys. Rev. A},
  volume = {112},
  issue = {4},
  pages = {042220},
  numpages = {13},
  year = {2025},
  month = {Oct},
  publisher = {American Physical Society},
  doi = {10.1103/fcjk-qbwh},
  url = {https://link.aps.org/doi/10.1103/fcjk-qbwh}
}

@misc{xiaoyue2024strategiesoptimizingdoublebracketquantum,
      title={Strategies for optimizing double-bracket quantum algorithms}, 
      author={Li Xiaoyue and Matteo Robbiati and Andrea Pasquale and Edoardo Pedicillo and Andrew Wright and Stefano Carrazza and Marek Gluza},
      year={2024},
      eprint={2408.07431},
      archivePrefix={arXiv},
      primaryClass={quant-ph},
      url={https://arxiv.org/abs/2408.07431}, 
}

@article{Gluza2024doublebracket,
  doi = {10.22331/q-2024-04-09-1316},
  url = {https://doi.org/10.22331/q-2024-04-09-1316},
  title = {Double-bracket quantum algorithms for diagonalization},
  author = {Gluza, Marek},
  journal = {{Quantum}},
  issn = {2521-327X},
  publisher = {{Verein zur F{\"{o}}rderung des Open Access Publizierens in den Quantenwissenschaften}},
  volume = {8},
  pages = {1316},
  month = apr,
  year = {2024}
}

@misc{mcmahon2025equatingquantumimaginarytime,
      title={Equating quantum imaginary time evolution, {R}iemannian gradient flows, and stochastic implementations}, 
      author={Nathan A. McMahon and Mahum Pervez and Christian Arenz},
      year={2025},
      eprint={2504.06123},
      archivePrefix={arXiv},
      primaryClass={quant-ph},
      url={https://arxiv.org/abs/2504.06123}, 
}

@article{Bild23,
author = {Marius Bild  and Matteo Fadel  and Yu Yang  and Uwe von Lüpke  and Phillip Martin  and Alessandro Bruno  and Yiwen Chu },
title = {Schrödinger cat states of a 16-microgram mechanical oscillator},
journal = {Science},
volume = {380},
number = {6642},
pages = {274-278},
year = {2023},
doi = {10.1126/science.adf7553},
URL = {https://www.science.org/doi/abs/10.1126/science.adf7553}
}

@article{vlastakis2013deterministically,
  title={Deterministically encoding quantum information using 100-photon {S}chr{\"o}dinger cat states},
  author={Vlastakis, Brian and Kirchmair, Gerhard and Leghtas, Zaki and Nigg, Simon E and Frunzio, Luigi and Girvin, Steven M and Mirrahimi, Mazyar and Devoret, Michel H and Schoelkopf, Robert J},
  journal={Science},
  volume={342},
  number={6158},
  pages={607--610},
  year={2013},
  publisher={American Association for the Advancement of Science},
doi = {https://doi.org/10.1126/science.1243289}
}

@article{he2023fast,
  title={Fast generation of {S}chr{\"o}dinger cat states using a {K}err-tunable superconducting resonator},
  author={He, XL and Lu, Yong and Bao, DQ and Xue, Hang and Jiang, WB and Wang, Zhen and Roudsari, AF and Delsing, Per and Tsai, JS and Lin, ZR},
  journal={Nature communications},
  volume={14},
  number={1},
  pages={6358},
  year={2023},
  publisher={Nature Publishing Group UK London},
doi = {10.1038/s41467-023-42057-0}
}

@article{wang2016schrodinger,
  title={A {S}chr{\"o}dinger cat living in two boxes},
  author={Wang, Chen and Gao, Yvonne Y and Reinhold, Philip and Heeres, Reinier W and Ofek, Nissim and Chou, Kevin and Axline, Christopher and Reagor, Matthew and Blumoff, Jacob and Sliwa, KM and others},
  journal={Science},
  volume={352},
  number={6289},
  pages={1087--1091},
  year={2016},
  publisher={American Association for the Advancement of Science},
doi = {10.1126/science.aaf2941}
}

@article{leghtas2015confining,
  title={Confining the state of light to a quantum manifold by engineered two-photon loss},
  author={Leghtas, Zaki and Touzard, Steven and Pop, Ioan M and Kou, Angela and Vlastakis, Brian and Petrenko, Andrei and Sliwa, Katrina M and Narla, Anirudh and Shankar, Shyam and Hatridge, Michael J and others},
  journal={Science},
  volume={347},
  number={6224},
  pages={853--857},
  year={2015},
  publisher={American Association for the Advancement of Science},
doi = {10.1126/science.aaa2085}
}

@article{yu2025schrodinger,
  title={Schr{\"o}dinger cat states of a nuclear spin qudit in silicon},
  author={Yu, Xi and Wilhelm, Benjamin and Holmes, Danielle and Vaartjes, Arjen and Schwienbacher, Daniel and Nurizzo, Martin and Kringh{\o}j, Anders and Blankenstein, Mark R van and Jakob, Alexander M and Gupta, Pragati and others},
  journal={Nature Physics},
  volume={21},
  number={3},
  pages={362--367},
  year={2025},
  publisher={Nature Publishing Group UK London},
doi = {https://doi.org/10.1038/s41567-024-02745-0}}

@article{goto2022probing,
  title = {Probing chaos by magic monotones},
  author = {Goto, Kanato and Nosaka, Tomoki and Nozaki, Masahiro},
  journal = {Phys. Rev. D},
  volume = {106},
  issue = {12},
  pages = {126009},
  numpages = {26},
  year = {2022},
  month = {Dec},
  publisher = {American Physical Society},
  doi = {10.1103/PhysRevD.106.126009},
  url = {https://link.aps.org/doi/10.1103/PhysRevD.106.126009}
}

@article{Bloch1997,
    author={Bloch, Anthony M. and Brockett, Roger W. and Crouch, Peter E.},
    title={Double Bracket Equations and Geodesic Flows on Symmetric Spaces},
    journal={Communications in Mathematical Physics},
    year={1997},
    month={Aug},
    day={01},
    volume={187},
    number={2},
    pages={357-373},
    issn={1432-0916},
    doi={10.1007/s002200050140},
    url={https://doi.org/10.1007/s002200050140}
}

@article{Mittnenzweig2017,
    author={Mittnenzweig, Markus and Mielke, Alexander},
    title={An Entropic Gradient Structure for {L}indblad Equations and Couplings of Quantum Systems to Macroscopic Models},
    journal={Journal of Statistical Physics},
    year={2017},
    month={Apr},
    day={01},
    volume={167},
    number={2},
    pages={205-233},
    issn={1572-9613},
    doi={10.1007/s10955-017-1756-4},
    url={https://doi.org/10.1007/s10955-017-1756-4}
}

@article{carlen2017,
    title = {Gradient flow and entropy inequalities for quantum {M}arkov semigroups with detailed balance},
    author = {Eric A. Carlen and Jan Maas},
    journal = {Journal of Functional Analysis},
    volume = {273},
    number = {5},
    pages = {1810-1869},
    year = {2017},
    issn = {0022-1236},
    doi = {https://doi.org/10.1016/j.jfa.2017.05.003},
    url = {https://www.sciencedirect.com/science/article/pii/S0022123617301878}
}

@article{cao2019gradient,
    author = {Cao, Yu and Lu, Jianfeng and Lu, Yulong},
    title = {Gradient flow structure and exponential decay of the sandwiched Rényi divergence for primitive {L}indblad equations with GNS-detailed balance},
    journal = {Journal of Mathematical Physics},
    volume = {60},
    number = {5},
    pages = {052202},
    year = {2019},
    month = {05},
    issn = {0022-2488},
    doi = {10.1063/1.5083065},
    url = {https://doi.org/10.1063/1.5083065}
}

@article{carlen2020,
author={Carlen, Eric A.
and Maas, Jan},
title={Non-commutative Calculus, Optimal Transport and Functional Inequalities in Dissipative Quantum Systems},
journal={Journal of Statistical Physics},
year={2020},
month={Jan},
day={01},
volume={178},
number={2},
pages={319-378},
issn={1572-9613},
doi={10.1007/s10955-019-02434-w},
url={https://doi.org/10.1007/s10955-019-02434-w}
}

@article{braunstein1994statistical,
  title = {Statistical distance and the geometry of quantum states},
  author = {Braunstein, Samuel L. and Caves, Carlton M.},
  journal = {Phys. Rev. Lett.},
  volume = {72},
  issue = {22},
  pages = {3439--3443},
  numpages = {0},
  year = {1994},
  month = {May},
  publisher = {American Physical Society},
  doi = {10.1103/PhysRevLett.72.3439},
  url = {https://link.aps.org/doi/10.1103/PhysRevLett.72.3439}
}

\end{document}